\documentclass[nohyper,notoc]{article} 
\usepackage{axodraw}
\usepackage{epsfig}

\usepackage[english]{babel}

\def\beq{\begin{equation}}
\def\eeq{\end{equation}}
\def\bea{\begin{eqnarray}}
\def\eea{\end{eqnarray}}
\def \lsim{\mathrel{\vcenter
     {\hbox{$<$}\nointerlineskip\hbox{$\sim$}}}}
\def \gsim{\mathrel{\vcenter
     {\hbox{$>$}\nointerlineskip\hbox{$\sim$}}}}
\def\gappeq{\mathrel{\rlap {\raise.5ex\hbox{$>$}}
{\lower.5ex\hbox{$\sim$}}}}
\def\lappeq{\mathrel{\rlap{\raise.5ex\hbox{$<$}}
{\lower.5ex\hbox{$\sim$}}}}

\begin{document}
\renewcommand{\thefootnote}{\fnsymbol{footnote}}

\vskip 5pt
\begin{center}
{\Large {\bf 
 Constraints on  two-lepton, two quark operators}}
\vskip 25pt
{\bf  Michael Carpentier  $^{1}$
and  Sacha Davidson $^{1,}$\footnote{E-mail address:
s.davidson@ipnl.in2p3.fr} 
} 
 
\vskip 10pt  
$^1${\it IPNL, Universit\'e Lyon 1, Universit\'e de Lyon, CNRS/IN2P3, 4 rue E. Fermi 69622 Villeurbanne cedex, France
}

\vskip 20pt
{\bf Abstract}

	\end{center}

\begin{quotation}
  {\noindent\small 
Physics from beyond the Standard Model, such as leptoquarks, 
 can  induce four fermion operators
involving a quark, an anti-quark, a lepton and an anti-lepton. We    
 update the (flavour-dependent)  constraints on the coefficients 
of such interactions, arising from collider searches for contact
interactions, meson decays and other rare processes. 
We then make naive estimates for the magnitude of the coefficients,
as could arise in texture models or from inverse hierarchies 
in  the kinetic term coefficients.
These ``expectations'' suggest that  rare Kaon decays
could be a good place to look for such operators.

\vskip 10pt
\noindent
}

\end{quotation}

\vskip 20pt  

\setcounter{footnote}{0}
\renewcommand{\thefootnote}{\arabic{footnote}}

\section{Introduction}

Various arguments, such as the ``hierarchy problem'' and the existance
of Dark Matter,  suggest New Physics (NP) with
 a   mass scale $m_{NP} \gsim G_{F}^{-1/2}$. 
The new interactions could manifest themselves, at energies below $m_{NP}$, 
as small deviations from   expected Standard Model rates,  or as  
new processes, absent in the Standard Model (SM). 
If $m_{NP}$ is accessible to the LHC, the new particles
could be discovered soon. 

To identify New Physics produced at colliders, it helps to have input
from low energy precision experiments. For instance, $G_F$  and
$\alpha_{em}$ were important inputs for the electroweak fit to LEP data. 
In  flavour physics, precision low
energy experiments searching for suppressed or
forbidden SM modes, are sensitive to New Physics
scales $\gg$ TeV. These bounds can be compatible
with TeV mass NP, provided there is some  ``similarity'' between  the
 flavoured interactions of the new particles and
of the SM (such as Minimal Flavour
Violation \cite{MFV,dAGIS}).  It is therefore 
interesting  to study the interplay
between  flavour and high energy
experiments, both current and future, as
was recently done in the  series of  workshops 
"Flavour in the era of the LHC"\cite{flavCERN}.

Contact interactions\cite{Buchmuller:1985jz}  involving 
two quarks and two leptons  are particularly relevant 
because the LHC can have $q\bar{q}$ in the initial state.
This paper  
compiles  bounds on the coefficients of  four fermion
interactions involving a quark, an anti-quark, a lepton
and an anti-lepton. These will be refered to as ``two quark,
two lepton'' interactions in the following. The
coefficients are assumed to be flavour-dependent,
 and the bounds are set assuming the presence
of one interaction, of given flavour indices, at a time.   
An update of these constraints  is timely, 
to take into account the improved limits
from  B factories. 

We consider  four fermion interactions which are  induced
by dimension six gauge invariant operators. 
The  bound on the coefficient of a  particular gauge
invariant operator can be obtained by consulting
the tables of bounds on all the four fermion
interactions it induces, and selecting the most
stringent one. We present bounds on the coefficients of 
(non gauge-invariant) four fermion interactions,
rather than  of gauge invariant operators, to allow
constraints to be extracted for an arbitrary basis
of gauge invariant operators. This is an artifact
of setting bounds by allowing one interaction at a time;
we could constrain
gauge invariant operators 
if we fit simultaneously to all the coefficients.

Various processes can constrain a given four fermion
interaction. We
attempt to find the best bound on each coefficient, and list it
in tables \ref{tab:eeuuND} - \ref{tab:CC}.
In the case of SM-allowed processes which are observed, 
the bound  is obtained by requiring that the SM-New Physics
interference contribute less than $ 2 \sigma$.  In
the case of unobserved decay rates, the New Physics is
bounded to contribute below the $90\% $ confidance level
experimental limit.  The bounds should be correct at the factor of
two level, but not to the two significant figures quoted. 

The limits compiled here are  mostly obtained from
observables to which the two quark two lepton interactions
contribute at tree level: 
contact interaction searches at colliders 
\cite{McFarland:1997wx,Schael:2006wu, Abe:1997gt},
 and various low energy processes  which are rare or suppressed in
the Standard Model : leptonic
and semi-leptonic  meson 
decays \cite{Amsler},  semi-leptonic $\tau$ decays \cite{Amsler}, 
 and $\mu -e$ conversion \cite{Dohmen:1993mp}.
The four fermion operators can also contribute in loops, for instance
one obtains a penguin by closing two legs and attaching a gauge boson. 
However, such  loop bounds  depend on the New Physics 
inducing the contact interaction,
as discussed in section \ref{sec:Zdec}. So we only
include estimated bounds from  $Z$ decay observables at LEP1 \cite{Zrefs},
because these were the only constraints we could find on
some higher generation operators.

Constraints have previously been compiled\footnote{For 
additional  pre-1995 references, see the references of
\cite{Davidson:1993qk}},
 on two quark and two lepton  contact interactions
 \cite{Barger:1997nf,Raidal:2008jk,Nelson:1997mr, Eichten:1983hw,  Gaillard:1974hs, Pich:1995vj,Ibarra:2004pe,Cashmore:1985xn,Buchmuller:1997hn,DiBartolomeo:1997mb, Cheung:2001wx,Zarnecki:1999je} as could arise in strongly coupled
models, and in other
 New Physics scenarios  such
as leptoquarks
 \cite{Buchmuller:1986zs,Buchmuller:1986iq,Davidson:1993qk, Herz:2002gq,Leurer:1993qx,Leurer:1993em, Blumlein:1997fm}
 and $Z'$s
 \cite{Cheung:2001wx,Cahn:1980kv,Salvioni:2009jp,He:2007iu,Rizzo}. 
As reviewed in section \ref{basis},
the most natural basis of  four fermion operators depends
somewhat on the New Physics scenario. Following the
 lepton chapter of the CERN
Workshop report  \cite{Raidal:2008jk} (which  gives bounds on contact
interactions involving  four leptons, and on  
some two lepton and two quark interactions),  we use operators where
the leptons and quarks are contracted, in most cases, among
themselves.

Meson factories, such as Belle and 
 NA62 \cite{na62},   will be studying rare
 decays with great precision in the upcoming years. 
They could search, for instance,  for
lepton flavour changing but generation diagonal
decays such as  
 $K^+ \to \pi^+ \mu^\pm e^\mp$
or  $B^+ \to K^+ \tau^\pm \mu^\mp$,
 which
  would be ``natural'' for leptoquarks
with generation diagonal couplings.
It would be useful to have some phenomenologically
motivated ``expectations'' for whether NP should
appear in second or third generation rare decays.  
So in the last section of this paper, we estimate
 the coefficients of the two-quark, two-lepton operators
in the Cheng-Sher ans\"atz \cite{Cheng:1987rs}
for dimensionless flavoured coupling constants, 
$\lambda^{ij}\sim \sqrt{y_i y_j}$, where
$y_i, y_j$ are the Yukawa couplings of
SM fermions $i$ and $j$. 
This  predicts the relative
rates of different processes,
which  allows to compare the
sensitivity  of different
processes to the underlying New Physics
mass scale.

\section{Operator Basis in the Effective Lagrangian}
\label{basis}

There are a large number of gauge invariant two quark two lepton operators at
dimension six,  not all of which are independent from each
other: equations of motion give relations between operators,
and Fiertz transformations  rearrange the
Lorentz contractions. In this section, 
 we review the four fermion operator
basis of \cite{Raidal:2008jk}, and discuss the
relation to the (non gauge invariant) 
four fermion interactions on which we quote bounds.

If  
 New Physics  is present  at the scale $m_{NP}$,
it can be  
added in the  SM as terms
in an  effective Lagrangian $\mathcal{L}^{SM}_{eff}$, 
which can be written,
at energies below $m_{NP}$ as  
an expansion in $1/m_{NP}$:
	\begin{equation}
	\mathcal{L}^{SM}_{eff}=\mathcal{L}_{0}+\frac{1}{m_{NP}}\mathcal{L}_{1}+
\frac{1}{m^{2}_{NP}}\mathcal{L}_{2}+\frac{1}{m_{NP}^{3}}\mathcal{L}_{3}+\cdots
	\label{1}
	\end{equation}
	where $\mathcal{L}_{0}$ is the renormalizable 
SM Lagrangian, and 
the $\mathcal{L}_{n}$   contain dimension $d=4+n$ 
operators, constructed out of Standard Model
fields,  and 
 invariant under the SM  $SU(3)\times SU(2)\times U(1)$
gauge symmetry.

The  SM  fermions are written as:
\begin{equation}
q_{i}=\left( 
\begin{array}{c}
u_{Li}\\
d_{Li}
\end{array}
\right) ~~,~
\ell_{i}=
\left( 
\begin{array}{c}
\nu_{Li} \\
e_{Li}
\end{array}
\right) 
 u_{Ri},\ d_{Ri},\ e_{Ri}
\label{2}	
\end{equation}
	where $i$ is the family/generation index.
(Notice that, in what follows, 
$\ell$ stands for the lepton doublet,  and $l$ for any
charged  lepton.)
Then  the basis of operators  used in  \cite{Raidal:2008jk} 
for the dimension six ${\cal L}_2$ contains
the   V$\pm$A  
 two lepton-two quark  operators:
\bea
	 \mathcal{O}^{ijkn}_{(1)\ell q}&=&
\left(\bar{\ell}_{i}\gamma^{\mu}P_L\ell_{j}\right)
\left(\bar{q}_{k}\gamma_{\mu}P_Lq_{n}\right), \label{O1lq}\\
&=&\left(\bar{\nu}_{i}\gamma^{\mu}\nu_{j} + \bar{e}_{i}\gamma^{\mu}P_L e_{j}\right)
\left(\bar{u}_{k}\gamma_{\mu}P_L u _{n} + \bar{d}_{k}\gamma_{\mu} P_L d_{n}
\right), \label{O1lqcomp}\\
	 \mathcal{O}^{ijkn}_{(3)\ell q}&=&
\left(\bar{\ell}_{i}\tau^{I}\gamma^{\mu}P_L\ell_{j}\right)
\left(\bar{q}_{k}\tau^{I}\gamma_{\mu}P_Lq_{n}\right),\nonumber \\
	 \mathcal{O}^{ijkn}_{e q}&=&
\left(\bar{e}_{i}\gamma^{\mu}P_Re_{j}\right)
\left(\bar{q}_{k}\gamma_{\mu}P_Lq_{n}\right),  \nonumber \\
	 \mathcal{O}^{ijkn}_{ed}&=&\left(\bar{e}_{i}\gamma^{\mu}P_{R}e_{j}\right)
\left(\bar{d}_{k}\gamma_{\mu}P_{R}d_{n}\right), \nonumber \\
	 \mathcal{O}^{ijkn}_{eu}&=&\left(\bar{e}_{i}\gamma^{\mu}P_{R}e_{j}\right)
\left(\bar{u}_{k}\gamma_{\mu}P_{R}u_{n}\right), \nonumber \\
	 \mathcal{O}^{ijkn}_{\ell u} 
&= & \left(\bar{\ell}_{i}\gamma^{\mu}P_L\ell_{j}\right)
\left(\bar{u}_{k}\gamma_{\mu}P_Ru_{n}\right) 
 \nonumber \\
	\mathcal{O}^{ijkn}_{ld} 
&=&  \left(\bar{\ell}_{i}\gamma^{\mu}P_L\ell_{j}\right)
\left(\bar{d}_{k}\gamma_{\mu}P_Rd_{n}\right) 
\label{4}	
\eea    
where the Lorentz, colour and SU(2) are  
contractions  are implicit within the parentheses, $\tau ^I$ are the Pauli
matrices ($\tau^3 = diag \{ 1, -1\}$),  and
  generation indices  are subscripts.
With the same notation (but now SU(2)  contraction
between the parentheses),  the  S$\pm$P  operators are:
\bea
	& \mathcal{O}^{ijkn}_{\ell qS}=\left(\bar{\ell}_{i}{e}_{j}\right)\left(\bar{q}_{k}u_{n}\right), \nonumber\\
	& \mathcal{O}^{ijkn}_{qde}=\left(\bar{\ell}_{i}e_{j}\right)\left(\bar{d}_{k}q_{n}\right).
\label{5}
\eea
See  \cite{Raidal:2008jk} for the expansion of these operators
on the component fields which participate in observable
interactions ({\it e.g.} $\nu_L$ and $e_L$ rather than $\ell$).
A similar expansion, for the four fermion operators induced
by  scalar leptoquarks, can be found in  \cite{DGI}.

After Electroweak Symmetry Breaking, an operator constructed
with electroweak doublets, for instance
${\cal O}_{(1) \ell q}$ of eqn (\ref{O1lq}), will induce four fermion
interactions among the doublet components (see eqn (\ref{O1lqcomp})). 
The data bound the coefficients of these four fermion interaction
among chiral  fields ($u_L, d_R, e_L $, etc),
so in our tables,
we quote bounds on the coefficients of four-fermion
interactions.
The best bound on the coefficient of a gauge invariant operator
will be the most restrictive limit to
be found on all the  four fermion interactions it induces. 

We quote bounds on the coefficients of  four fermion 
interactions, rather than
those of gauge invariant operators, to minimise the 
dependence on operator basis choice.  To motivate this,
consider the alternative gauge invariant operator ${\cal O}_{alt}$,
that could be induced by a scalar leptoquark:
\bea
{\cal O}_{alt}& =& 
 ( \overline{q^c}  \varepsilon \ell)(\bar{\ell} \varepsilon q^c) 
\label{Alt} \\ &=&
\frac{1}{2} \left[ (  \overline{u}  \gamma^{\mu}P_L u )(\bar{e}
\gamma_{\mu}P_L e)
+
 (  \bar{d}  \gamma^{\mu} P_L u )(\bar{\nu}
\gamma_{\mu}  P_L e)
+
 (  \bar{u}  \gamma^{\mu} P_L d )(\bar{e}
\gamma_{\mu}  P_L \nu)
+ 
 ( \bar{d}  \gamma^{\mu}P_L  d)(\bar{\nu}
\gamma_{\mu} P_L \nu) \right] 
\nonumber
\eea
where $\varepsilon$ gives the antisymmetric SU(2) contraction.
Using the  identity $\delta_{ab} \delta_{cd} +  \sum_i [\sigma_i]_{ab} [\sigma_i] _{cd}
= 2 \delta_{ad} \delta_{bc}$ for the SU(2) contraction
between $q$ and $\varepsilon \ell$, the operator (\ref{Alt})
can be rewritten in terms of ${\cal O}_{(1) \ell q}$ and
 ${\cal O}_{(3) \ell q}$. However, since the
bounds are computed by allowing  the coefficient of
one operator at a time to be non-zero, the constraints
on the coefficients of ${\cal O}_{alt}$
cannot be obtained easily from the bounds on the coefficients
of  ${\cal O}_{(1) \ell q}$ and
 ${\cal O}_{(3) \ell q}$. We therefore quote
bounds on the more ``physical'' four fermion interactions,
which allows bounds on any choice of gauge invariant
dimension 6 operators to be extracted from our
tables.

We choose the factors of 2 in the normalisation of the operators  such
that the coefficient $ { C_{ijkn}}/{m_{NP}^2}$
replaces $-4 G_F/\sqrt{2}$ as the  coupling constant in the Feynman rules. 
So  the hermitian operators of eqn (\ref{4}) 
are included as
\beq
{\cal L}  = ... + \frac{1}{2}\sum_{i,j,k,n = 1}^3 \frac{ C_{ijkn}}{m_{NP}^2}
 \mathcal{O}^{ijkn}  + h.c.
\eeq
where the 1/2 compensates  the $+ h.c.$. The generation indices
of the  $S \pm P$ operators
also are summed  $1..3$, and there is a $+ h.c$ but no 1/2
because the operators are not hermitian.

In order to set bounds on dimensionless quantities, 
the coefficients of the two lepton-two quark operators in (\ref{4})
and (\ref{5}) are normalized relative to the Fermi interactions:
\bea
	 \frac{C^{ijkn}_{(n)\ell q}}{m^{2}_{NP}}=
-\frac{4G_{F}}{\sqrt{2}}\epsilon^{ijkn}_{(n)\ell q},
&&
	 \frac{C^{ijkn}_{e q}}{m^{2}_{NP}}=
-\frac{4G_{F}}{\sqrt{2}}\epsilon^{ijkn}_{e q}, \\
 \frac{C^{ijkn}_{eu}}{m^{2}_{NP}}
=-\frac{4G_{F}}{\sqrt{2}}\epsilon^{ijkn}_{eu}, &&  
 \frac{C^{ijkn}_{ed}}{m^{2}_{NP}}=
-\frac{4G_{F}}{\sqrt{2}}\epsilon^{ijkn}_{ed},\\
   \frac{C^{ijkn}_{\ell u}}{m^{2}_{NP}}=
\frac{4G_{F}}{\sqrt{2}}\epsilon^{ijkn}_{\ell u}, &&
    \frac{C^{ijkn}_{\ell d}}{m^{2}_{NP}}
=\frac{4G_{F}}{\sqrt{2}}\epsilon^{ijkn}_{\ell d}, \\ 
   \frac{C^{ijkn}_{\ell qS}}{m^{2}_{NP}}=
-\frac{4G_{F}}{\sqrt{2}}\epsilon^{ijkn}_{\ell qS}, &&
  \frac{C^{ijkn}_{qde}}{m^{2}_{NP}}=
-\frac{4G_{F}}{\sqrt{2}}\epsilon^{ijkn}_{qde}.      
\label{7}
\eea
Notice that the first two indices, $ij$,
 are always lepton indices, and the last two,
$kn$, are always quarks.

\section{Constraints from Rare Decays}

A multitude of  precision experiments searching for 
rare meson decays  provide stringent bounds on
the generation-changing interactions of New  Physics.
We use the experimental bounds compiled 
and  averaged by the Particle Data Group \cite{Amsler}. 
Constraints on New Physics have
been obtained from this data in the papers mentioned
in the introduction,  and  
 in more  recent years by
\cite{Dreiner:2006gu,Matsuzaki:2009fk,Campbell:2008um,Smirnov:2007hv}, 
including CP violation effects\cite{Saha:2010vw},
asymmetries in final state distributions
 \cite{Alok:2009tz}, 
and discrepancies between the data and the SM
 \cite{Dorsner:2009cu,Golowich:2007ka}.

\subsection{Leptonic meson decays}

In the presence of New Physics, 
the leptonic decay rate of a charged pseudoscalar   meson $M_{kn}$
(for instance $\pi^+$),  
made of constituent quarks
$\bar{q}_k q_n$ (for instance $\bar{d} u$), can be written:
\begin{eqnarray}
 \Gamma \left(M_{kn}\rightarrow l^{i}\bar{l}^{j}\right)
& =& \frac{kG_F^2}{\pi m^{2}_{M}} 
\Bigg\{ (\epsilon^{ijkn}_{S+P})^{2}\widetilde{P}^{2}
\left(m^{2}_{M}-m^{2}_{i}-m^{2}_{j}\right)+  \nonumber\\  
&& \left[ V_{kn}-\epsilon^{ijkn}_{V-A} \right]^{2}
\widetilde{A}^{2}\left[(m^{2}_{M}-m^{2}_{i}-m^{2}_{j})(m^{2}_{i}+m^{2}_{j})
+4m^{2}_{i}m^{2}_{j}\right]+  \nonumber\\ 
&& 2\left[{\epsilon^{ijkn}_{S+P}\epsilon^{ijkn}_{V-A}}
 -{ V_{kn}\epsilon^{ijkn}_{S+P}}\right]
\widetilde{A}\widetilde{P}m_{j}\left(m^{2}_{M}+m^{2}_{i}-m^{2}_{j}\right)\Bigg\} ~.
\label{8}
\end{eqnarray}
The  expectation values of quark currents 
 are  taken to have their current  algebra values:
$$\widetilde{A}P^\mu= \frac{1}{2} 
\langle 0| \overline{q} \gamma^\mu \gamma^5 q |M \rangle 
=  \frac{f_{M}P^\mu}{2} 
~~~~~~~
\widetilde{P}=  \frac{1}{2} 
\langle 0| \overline{q}  \gamma^5 q |M \rangle 
= \frac{f_{M}m_{M}}{2}\frac{m_{M}}{m_{k}+m_{n}}~~.$$ 
These formulae are used for pions, kaons, D 
\footnote{We extract bounds from $D_s^+$ decays, 
although there is a few $\sigma$ discrepancy among
these  decays and lattice results, which has
been discussed in leptoquark models
 \cite{Dorsner:2009cu}.}
  and B mesons,
with  meson decay constants
$f_M$  taken from  \cite{Amsler},  and from \cite{Bernard:2009wr}
for the $B$s (we take $f_{B0} = f_{B+} = 195$ MeV, and $f_{Bs} = 243$ MeV).
 $P^\mu$ is the momentum of the meson, and
$k$ is the magnitude of the lepton 3-momentum 
in the center-of-mass frame:
	\begin{equation}
k^{2}=\frac{1}{4m^{2}_{M}}\left[\left(m^{2}_{M}-\left(m_{i}+m_{j}\right)^{2}\right)\left(m^{2}_{M}-\left(m_{i}-m_{j}\right)^{2}\right)\right]
\label{9}
\end{equation}

In the Standard Model,  charged pseudoscalar mesons   decay to light leptons
via $s$-channel $W$ boson exchange. This  amplitude is
suppressed by the lepton mass, due to angular momentum conservation:
in the relativistic limit, where helicity $\simeq $ chirality,
the chirality of one of the left-handed leptons must be flipped.
This suppression can be seen in the bracket multiplying $\tilde{A}^2$
in eqn (\ref{8}) , and is precisely measured
in the $R_\pi$  \cite{Rpi} and $R_K$ ratios
\bea
R_{\pi} & \equiv& \frac{\Gamma (\pi^+ \to e^+ \nu)}
{\Gamma (\pi^+ \to \mu^+ \nu)} \nonumber \\
R_{\pi}{\Big |}_{theory} &
= & \frac{m_e^2}{m_\mu^2} 
\frac{(m_\pi^2 - m_e^2)^2}{(m_\pi^2 - m_\mu^2)^2} {\Big ( }
1 + \Delta {\Big )} = (1.2354 \pm 0.0002) \times 10^{-4}
\label{Rpi}
\eea
(where $\Delta$ are SM corrections to the given tree
formula). 
 New Physics that induces  the $S+P$ type
 operators of eqn (\ref{5}) does not suffer this suppression,
as can be seen from the first term of eqn  (\ref{8}), so is
tightly constrained by the decays $M_{kn} \rightarrow l^i \bar{l}^j$.

Bounds from a given process, on the  $\epsilon^{ijkn}$ coefficients
of two quark two lepton interactions,
 are obtained by allowing one
  $\epsilon^{ijkn}$   at a time to be
non-zero\footnote{ So the  
$\epsilon_{V-A} \epsilon_{S+P}$  interefence term, 
which is included to make the decay 
rate self-consistent, will not be used  to
set bounds. }. For instance, in the charged meson decays  which occur at
tree level in the SM,   we successively allow all
the operators involving a neutrino. When the neutrino flavour is
the same as in the SM amplitude,  these  
New Physics interactions can interfere with  the  SM,
giving a contribution to the decay rate $\propto V \epsilon$. 
For other neutrino flavours, the New Physics
contributes  $\propto \epsilon^2$.

We also  use eqn (\ref{8})  to describe the Flavour
Changing Neutral Current (FCNC) decays of
neutral mesons (such as $K_L
\to \bar{\mu} e$, or $B \to \mu \bar{\mu}$)
 induced by  the effective operators of eqn (\ref{4}) and
(\ref{5}).  The SM contribution in these decays
is very small,  so  to use  eqn 
(\ref{8}), the term proportional to 
 the CKM matrix element $V_{kn}$,
is set to zero.
These processes  give bounds on     the $V\pm A$
type interactions
$(\bar{e}_i \gamma^\mu P_{L,R} e_j) 
(\bar{u}_k \gamma_\mu P_{L,R} u_n)$ 
and
$(\bar{e}_i \gamma^\mu P_{L,R} e_j) 
(\bar{d}_k \gamma_\mu P_{L,R} d_n)$.

\subsection{Semi-leptonic decays}

Semi-leptonic decays of mesons and hadrons can occur
at tree level in the SM, via charged current interactions,
and the good agreement between theory and data can be
used to set bounds on New Physics in charged currents (see {\it e.g.}
\cite{CJGA} for flavour diagonal bounds). We will use the  $K^+$ form
factors and the observed  unitarity of
the  CKM matrix  to set
bounds on  charged current two quark, two lepton interactions. 
We start by discussing bounds from  generation changing,
``neutral current'' decays, which are
forbidden at tree level in the SM.

\subsubsection{Neutral current decays}

Semileptonic meson  decays can be used to
set bounds  on  generation changing neutral
current  operators of the  $V \pm A$ -type,  
by using various isospin symmetries to relate 
the New Physics operator to SM-induced charged
current processes. We use the following approximations:
\bea
\left\langle K^{+}\right|\widetilde{O}\left|\pi^{0}\right\rangle
&\approx &
\frac{1}{\sqrt{2}}\left\langle K^{+}\right|\widetilde{O}
\left|\pi^{+}\right\rangle \qquad  \\
 \left\langle D^{+}\right|\bar{u}\gamma^{\mu}Pc\left|\pi^{+}\right\rangle
& \approx&
 \langle D^{0} |\bar{d}\gamma^{\mu}Pc |
\pi^{-} \rangle \qquad \textrm {with $P=P_{L}$, $P_{R}$} \\
\left\langle B^{+}\right|\bar{d}\gamma^{\mu}\gamma^{5}b\left|\pi^{+}\right\rangle
&\approx&
\left\langle B^{+}\right|\bar{u}\gamma^{\mu}\gamma^{5}b\left|
\pi^{0}\right\rangle \qquad 	 \\
\langle B^{0} |\bar{u}\gamma^{\mu}\gamma^{5}b |\pi^{-} \rangle
&\approx&
\sqrt{2} \left\langle B^{+}\right|\bar{u}\gamma^{\mu}\gamma^{5}b\left|
\pi^{0}\right\rangle
\label{10}	
\eea
where $\tilde{O}$ is some isospin 1/2 operator.

	The constraints arise from generation changing decays of
$K$, $D$ or $B$ mesons,
such as $K^+ \to \pi^+ e_i^+ e_j^-$  or 
 $K^+ \to \pi^+ \nu_i \bar{\nu}_j$ \cite{Artamonov:2008qb}.
We estimate the decay rates  for such
processes,  induced by
the New Physics $V \pm A$ operator,  as
 the squared  ratio of the  NP to SM quark
matrix elements, times the measured rate
for a SM allowed process (for instance,
 $K^+ \to \pi^0 e^+ \nu$).
In the case of $K$ decays, this has errors due to lepton  mass effects. 
For  semi-leptonic $B$ decays, the errors are more significant. 
For instance, we approximate 
\beq
\frac{\Gamma(B^+ \to K^+ \tau^\pm \mu^\mp)}{
\Gamma(B^+ \to D^0 e^+ \nu)}
\simeq \frac{ |\epsilon^{\tau \mu b s}|^2}{|V_{cb}|^2} 
\eeq

\subsubsection{$K^+_{l3}$ and (pseudo)scalar operators}

Following \cite{Herz:2002gq}, the  decays
$K^{+}_{e 3} = K^+ \to \pi^0 e^+ \nu$ and $K^{+}_{\mu 3}$  can be 
used to constrain  the  $S \pm P$  operators ${\cal O}_{\ell q S}
= (\bar{\ell} e)(\bar{q} u)$ (see eqn (\ref{5})) \footnote{ Due to
the $u_R$, these operators are not strictly constrained by the
rare neutral kaon decays like $K_L \to \bar{\mu} e$.}.
The bound arises from the experimental limit 
\cite{Yushchenko:2004zs,Yushchenko:2003xz} on 
${f_{s}^{exp}}/{f_{+}(0)}$, where $f_{s}^{exp}$ is the 
scalar decay constant of the kaon,
 and $f_{+}$ one of the 
$V-A$ decay constants, both defined as follows:
\begin{equation}
\label{13}	
\mathcal{M}=G_{F}V_{us} \Bigg\{-(\bar{\nu_{l}}^{i}_{L}\gamma_{\mu}l^{j}_{L})
(f_{+}p^{\mu}+f_{-}q^{\mu})+2m_{K}
(\bar{\nu_{l}}^{i}_{L}l^{j}_{R})f^{exp}_{s}\Bigg\} 
\end{equation}
The  matrix element induced by the SM and $S\pm P$ New Physics, is:
\begin{equation}	
\mathcal{M}= -G_{F}V_{su}(\bar{\nu_{l}}^{i}_{L}\gamma_{\mu}l^{j}_{L})(f_{+}p^{\mu}+f_{-}q^{\mu})+
\frac{1}{\sqrt{2}}\frac{C^{ijkn}_{\ell qS}}{m^{2}_{Lq}}m_{K}f^{NP}_{s}(\bar{\nu_{l}}^{i}_{L}l^{j}_{R})
\label{14}
\end{equation}
with $p^{\mu}=(p_{K}+p_{\pi})^{\mu}$ and $q^{\mu}=(p_{K}-p_{\pi})^{\mu}$. This
implies
\begin{equation}
f^{exp}_{s}=\frac{C^{ijkn}_{\ell qS}}{m^{2}_{Lq}}
\frac{\sqrt{2}}{4G_{F}V_{su}}f^{NP}_{s}
\label{15}
\end{equation}
With the  current algebra identity:
\begin{equation}
f^{NP}_{s}=\frac{1}{2m_{K}}\frac{m^{2}_{K}-m^{2}_{\pi}}
{m_{s}-m_{u}}f_{+}+\frac{q^{2}}{(m_{s}-m_{u})2m_{K}}f_{-}
\label{16}
\end{equation}
we obtain
\begin{equation}
\frac{C^{ijkn}_{\ell qS}}{m^{2}_{Lq}}\frac{\sqrt{2}}{4G_{F}V_{su}}\frac{1}{2m_{K}}\frac{m^{2}_{K}-m^{2}_{\pi}}{m_{s}-m_{u}}\leq\left[\frac{f_{s}}{f_{+}(q^{2}=0)}\right]_{exp}
\label{17}
\end{equation}

\subsubsection{CKM Unitarity}

We can obtain constraints on the coefficient  $\epsilon^{ijkn}$ 
of the charged current four
fermion interaction $(\bar{\nu}_i \gamma^\mu P_L e_j)$
$(\bar{d}^k \gamma_\mu P_{L} u)$, from the observed unitarity
of the CKM matrix $V$. We allow   $\epsilon^{ijkn}$
for only one combination $ijkn$,  and impose that 
CKM remain unitary within the $2 \sigma$ uncertainties\cite{Amsler}
on its elements. So for instance, $ \epsilon^{\nu_i e  d u}$
would contribute to nuclear $\beta$ decay
like a shift  $V_{ud} \to
V_{ud} + \epsilon^{\nu_i e  d u}$. So we obtain
\beq
1-|V_{ud}|^2 - |V_{us}|^2 = 2 V_{ud} \epsilon^{\nu_e e  d u}
+| \epsilon^{\nu_i e  d u}|^2 \pm 4V_{us} \sigma_{us}
\label{CKMU}
\eeq
where  $ V_{us}\pm  \sigma_{us} = 0.2255 \pm 0.0019$, and    we neglect the 
experimental uncertainty on $V_{ud} = 0.97418 \pm .00027$. The
left hand side of this equality is $\simeq 10^{-4}$,
which gives the bounds quoted in the tables. 
Comparing $\tau^- \to \pi^- \nu$ to $\pi \to \mu \bar\nu$
 determines the ratio of the
leptonic couplings to the $W$: $g_\tau /g_\mu = 0.996 \pm.005$\cite{Pich},
which we translate to  $2 \epsilon^{\nu_\tau \tau  d u},$
$|\epsilon^{\nu_j \tau  d u}|^2 <2 \times .01$. 
Unitarity of the first row of $V$ also
gives    $2 V_{us} \epsilon^{\nu_l l  s u} ,
|\epsilon^{\nu_j l  s u}|^2 < 4  V_{us} \sigma_{us}$,
where  $j \neq l$, and $ l = e$ or $\mu$. For
$l = \tau$, bounds on  $\epsilon^{\nu_j \tau  s u}$ can be obtained from the
strange decays of $\tau$s, which  give $V_{us}$ with an
uncertainty \cite{tauVus,Amsler} $\sigma_{us} = 0.0027$. 
Similarly, unitarity of the first colomn of $V$, 
gives   $2 V_{cd} \epsilon^{\nu_l l  d c} ,
|\epsilon^{\nu_j l  d c}|^2 < 4  V_{cd} \sigma_{cd}$,
where  $j \neq l$, and $ l = e$ or $\mu$. $V_{cd}$
can be determined from $D \to \pi l \nu$ decays
($\sigma_{cd} = 0.024$), and is most accurately
determined in $\nu_\mu$ scattering ($\sigma_{cd} = 0.011$)
\cite{Amsler}.
The small mass difference $m_D - m_\tau$ makes 
$\epsilon^{\nu_j \tau  d c}$ less easy to constrain; by
comparing the upper bound\cite{Amsler}  
$BR(D^- \to \tau \bar{\nu}) < 1.2\times 10^{-3}$
and  $BR(D^- \to \mu \bar{\nu}) = 3.38\times 10^{-4}$ to
eqn (\ref{Rpi}), one can require
$\epsilon^{\nu_j \tau  d c} < V_{cd}$.
Finally, unitarity also implies that contributions 
of $\epsilon^{\nu_j l  s c}$   should  fit in
the uncertainty in  $V_{cd} = 1.04 \pm .06$, which
is determined from decays   with $\ell  = e,\mu$ and $\tau$.
Since the uncertainty is from  the lattice QCD determination
of the form factor,  we use it for all lepton generations.

It is more difficult to use unitarity to constrain the
charged current interactions involving $cb$,  and $ub$.
Instead, we  require  $\epsilon^{\nu_j l_i xb} \lsim V_{xb}$,
for $x = c$ and $u$, and all charged leptons  $l_i$ (since
$V_{xb}$ are measured in decays to all lepton flavours).

\subsection{Tau decays}
	
	Constraints on New Physics 
from rare tau decays, including
loop processes,   have been studied in \cite{Gabrielli1,Pich:1995vj,Kanemura:2005hr}.
Here we consider tree level $\tau$ decays, such
 as the lepton flavour violating  
 $\tau \rightarrow lM$,  where $l$ is $e$ or $\mu$ 
and M is a meson lighter than the 
tau ($\pi$ and K). 
The quark matrix elements and kinematic
factors are estimated 
by assuming that the
 decay rate was  the appropriate $|\epsilon|^2
\times$ the measured charged current SM process. 
For instance, a constraint can be obtained from
the experimental upper bound on $\tau \to \mu \pi^0$, by
comparing it to $\tau \to \nu_\tau \pi^-$. 
This means that the mass of final state leptons is
neglected, and the matrix elements are assumed to
satisfy
 \cite{Davidson:1993qk}:
\bea
& \left\langle 0\right|\bar{u}\gamma^{\mu}\gamma^{5}u\left|\pi^{0}\right\rangle\approx\left\langle 0\right|\bar{u}\gamma^{\mu}\gamma^{5}d\left|\pi^{-}\right\rangle \\
& \left\langle 0\right|\bar{d}\gamma^{\mu}\gamma^{5}s\left|K^{0}_{s}\right\rangle\approx\left\langle 0\right|\bar{u}\gamma^{\mu}\gamma^{5}s\left|K^{-}\right\rangle
\label{19}
\eea

\section{Other Processes}

\subsection{Contact Interactions and Atomic Parity Violation}
\label{contint}

The non-observation of ``contact interactions'' at colliders implies
bounds on flavour diagonal four fermion operators.
The constraints
 are  quoted as lower bounds on
a scale $\Lambda$ (in TeV), which,
in the case where two of the fermions
are electrons, is  defined via the Lagrangian
\begin{equation}
\mathcal{L}=\frac{4 \pi}{\Lambda^{2}}
\left[\eta_{LL}(\bar{e}\gamma_{\mu} P_L e )(\bar{\psi}\gamma^{\mu}P_L \psi)
+\eta_{RR}(\bar{e}\gamma_{\mu}P_R e)(\bar{\psi}\gamma^{\mu}P_R \psi)
+\eta_{LR}(\bar{e}\gamma_{\mu}P_L e)(\bar{\psi}\gamma^{\mu}P_R \psi)
+\eta_{RL}(\bar{e}\gamma_{\mu}P_R e)(\bar{\psi}\gamma^{\mu}P_L \psi)
\right]
\label{28}
\end{equation}	
where  the $\eta$s are $\pm 1$ or 0, and $\psi$ can be
any of $\{ \mu, \tau, u , d, s, c, b\}$.
Bounds are quoted on the $\Lambda$s
with the notation
\footnote{We here follow the notation of 
\cite{Schael:2006wu}, who put the electron current before the
quark in the four fermion interaction; notice that the Tevatrom
experiments invert the order, so that $\Lambda_{LR}$ from LEP
is $\Lambda_{RL}$ from the Tevatron.}
$ \Lambda^{\pm}_{LL} \equiv  \Lambda$  for $  (\eta_{LL},\ \eta_{RR}, \ \eta_{LR} ,  
\eta_{RL})
 = (\pm1, 0, 0,0 )$.

The 
CCFR neutrino experiment
 \cite{ McFarland:1997wx}, which scattered
mostly muon neutrinos off nuclei,
sets bounds on operators
of the form  $(\bar{\nu}_\mu \gamma \nu_\mu)(\bar{q}_1 \gamma q_1)$,
for $q_1$ a first generation quark.
The most recent bounds  that we found 
 on operators
of the form  $(\bar{\mu} \gamma \mu)(\bar{q}_1 \gamma q_1)$,
are from D0 \cite{D0data}.
For  operators of the form
 $(\bar{e} \gamma  e)(\bar{s}\gamma s)$ 
  and   $(\bar{e} \gamma e)(\bar{c} \gamma c)$,
we use the   bounds from LEP 
 \cite{Schael:2006wu}, and
for first generation
quark 
  operators 
 $(\bar{e} \gamma  e)(\bar{q}_1\gamma q_1)$
we follow  \cite{Cheung:2001wx}  in combining the   bounds from LEP 
 \cite{Schael:2006wu}, 
the
 Tevatron \cite{D0data}  HERA
\cite{HERAdata}  and Atomic Parity
Violation.
The LEP bounds assume the   contact interaction
is simultaneously present for all the light quarks
$\{ u, d, s \}$.
Since we wish to constrain
the four fermion interactions ``one
quark flavour at a time'',  we divide the LEP $\Lambda$s by $\sqrt{3}$ to
obtain our bounds. 
The D0 \cite{D0data} and HERA
\cite{HERAdata}
bounds 
assume the same contact interaction of
the leptons with $u$s and $d$s.  So for up quarks,
we use the given
bounds on $\Lambda$, and for $d_R$ quarks, we estimate
$\Lambda \sim \frac{1}{3}\Lambda_{D0,HERA}$, 
and $\Lambda \sim \frac{1}{\sqrt{3}}\Lambda_{D0,HERA}$ for $d_L$ quarks.

Atomic Parity Violation (APV) experiments\cite{Young:2007zs,APV,Amsler}
 measure the weak charge
of nuclei:
\beq 
Q_W = -2 {\Big [} C_{1}^{eu}(2Z + N) +  C_1^{ed}(Z + 2N) 
{\Big ]}
\eeq
where  $C_1^{eq}$s  is the dimensionless
coefficient of  the chiral four fermion
interaction $- \sqrt{2} G_F(\bar{e} \gamma^\mu P_{L,R} e)
(\bar{q} \gamma^\mu  q)$, so can be
replaced by any  of our $\epsilon$s.
Experiments with cesium \cite{APV}  
agree with the SM to within $\sim 1 \sigma$,
and give  a data- theory uncertainty \cite{Amsler} of  
$\Delta( C_1^{eu} +  C_1^{ed}) \simeq 0.0008 \pm 0.0013$,   
$\Delta( C_1^{eu} -  C_1^{ed}) \simeq 0.0015 \pm 0.0015$.   
Since we wish to set bounds separately
on  the coefficients of four fermion interactions
involving $u$s or $d$s, we  take the (2$\sigma$)
bound from APV to be $|\epsilon| < .03$.

As there are several experiments which set 
bounds of similar magnitude 
on some of the  first generation contact interactions, 
the combined bound is more restrictive. Such a combination
was performed several years ago by Cheung\cite{Cheung:2001wx},
and despite the improvements in the data, the bounds
of  \cite{Cheung:2001wx} are still the most restrictive
and are quoted in the PDB. 
 However, 
we do not use them,  because  they 
apply to the coefficients of a 
particular choice  of  gauge invariant operators
(the same as eqn (\ref{4})),
and in some cases,
the contact interactions were assumed to have the
same strength  for  the first and second generations
\footnote{In the case where the coefficients
of the operators are  flavour independent,
constraints have been calculated
more recently in \cite{Han:2004az}. This
analysis allows the coefficients of 21
dimension six operators to be simultaneously
non-zero.}.
Instead,  for $u$ and $d$ quarks separately,
we  combine the  limits
from the four experiments   as
\beq
\epsilon_{com} = \frac{1}{w} \sum_x \frac{\epsilon_x}{\sigma^2_x}
\pm \sqrt{\frac{1}{w}}
\eeq
where $\epsilon_x \pm \sigma_x$ is the
$95\%$ CL range allowed by  an
experiment, and $w = \sum_x 1/\sigma^2_x$. The central
values $\epsilon_{com}$ are $< \sqrt{1/w} \sim .01$, so 
 they are  all approximated as  zero. The resulting
limits  are 
listed in the tables.

\subsection{Decays of the Z}
\label{sec:Zdec}

 New Physics which induces
two lepton two quark  operators  at tree
level, can also mediate penguin/dipole operators
at one loop. 
In general, we avoid bounds from processes where 
the New Physics  contributes in loops,
because  they  are more  sensitive to the New
Physics details. 
However, in this section, we estimate bounds
from the $Z$ vertex, because these are
the only constraints we could find on certain
higher generation operators.

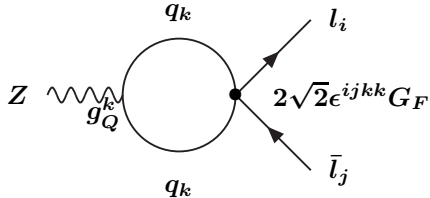
\begin{figure}[ht]
\unitlength.5mm
\SetScale{1.418}
\begin{boldmath}
\begin{center}
\begin{picture}(80,60)(0,0)
\ArrowLine(50,20)(70,40)
\ArrowLine(70,00)(50,20)
\CArc(35,20)(15,0,360)
\Text(-5,20)[r]{$Z$}
\Text(75,40)[l]{$l_i$}
\Text(75, 0)[l]{$\bar{l}_j$}
\Text(15,15)[c]{$g_Q^k$}
\Photon(00,20)(20,20){2}{4}
\Text(50,20)[c]{$\bullet$}
\Text(60,20)[l]{$2 \sqrt{2} \epsilon^{ijkk} G_F$}
\Text(35,42)[c]{$q_k$}
\Text(35,-5)[c]{$q_k$}
\end{picture}
\end{center}
\end{boldmath}
\caption{
One  loop  contribution 
  to   $ Z \to l_i \bar{l}_k$ from
an effective vertex $2 \sqrt{2} \epsilon^{ijkk} G_F$
 $(\bar{q}_k \gamma^\alpha P_Q q_k)
(\bar{l}_i \gamma_\alpha  P_Z l_j)$. $l_i$ are charged
leptons.  Combined with the
LEP1 Z data, such a diagram can set  bounds
on operators involving two top quarks, and/or two $\tau$s,
and /or charged leptons of different flavour.}
\label{Zdec}
\end{figure}

We estimate the
contribution of the contact interactions to  the
two-fermion-$Z$ boson operators
by closing   two fermion legs, and attaching a gauge boson
(see figure \ref{Zdec}). Notice however, that  this contact
interaction part may not be the full  contribution;
in the case of leptoquarks \cite{Zrefs}, there are two diagrams in
the full theory,
with the gauge boson attached to either the leptoquark
or the fermion. Only the fermion diagram
is included in the effective theory estimate
we will make here. 
We make these estimates for the
$Z$ vertex, rather than the photon, 
because closing the quark  legs of 
 the $V \pm A$ operators of eqns (\ref{4}),
  and attaching a photon, 
 gives no contribution to the lepton-lepton-photon dipole.
This is an artifact of our effective
field theory approach and operator basis choice,
since
there can be  bounds \cite{Benbrik:2008si,Gabrielli1} from dipole operators
($\mu \to e \gamma$, $b \to s \gamma$) on
leptoquarks and $Z'$s.

Four fermion operators of the form $(\overline{q}_k \gamma^\alpha P_Q  q_k)
(\overline{l}_i \gamma_\alpha P_Z l_j)$ , where $l_i$ and $l_j$
are charged leptons, can contribute at one-loop to
the decays $Z \to \bar{l}_i l_j$ via the diagram of figure \ref{Zdec}.
For $i = j$, the effective operator would interfere with
the SM amplitude, and  contribute linearly  to \cite{Amsler}
\beq
R_{l_j} = \frac{\Gamma(Z \to hadrons)}{\Gamma(Z \to l_j \bar{l}_j)} = 
\left\{
\begin{array}{ll}
  20.804
 \pm .050 
&  l = e \\
  20.785
 \pm .033 
&  l = \mu \\
  20.764
 \pm .045  
&  l = \tau
\end{array}
\right. 
\eeq
For $i \neq j$, the loop amplitude squared would  induce 
lepton flavour changing $Z$ decays,  whose branching ratios are
bounded 
\beq
BR(Z \to l_i \bar{l}_j) <
\left\{
\begin{array}{ll}
1.7 \times 10^{-6} & ij = e \mu\\
9.8 \times 10^{-6} & ij = e \tau\\
1.2 \times 10^{-5} & ij = \mu \tau\\
\end{array}
\right. 
\eeq
These  data  gives bounds of order $\epsilon \lsim 1$, which can be
 interesting for  higher generation quarks and/or leptons.

If the $Z$ coupling to fermions  $f_j$ is written
\beq
{\cal L} = - \frac{e}{2 s_W c_W} \overline{f}_j \gamma^\rho (g_L^f P_L
+ g_R^f P_R) f_j Z_\rho
\eeq
then the loop of figure \ref{Zdec} modifies the couplings $g_Z^j$
by adding a contribution 
$\delta g_Z^{ij} $, where the subscript $Z = L,R$ 
is the chirality of the final state
leptons.  
 Formulae for the vacuum polarisation loop 
can be found, for instance, in  chapter 21 of \cite{Peskin:1995ev}.
An operator  $2 \sqrt{2} \epsilon^{ijkk} G_F$
 $(\bar{q}_k \gamma^\alpha P_Q q_k)
(\bar{l}_i \gamma_\alpha  P_Z l_j)$ gives
\beq
\delta g_Z^{ij} \simeq \frac{N_c  2 \sqrt{2} \epsilon^{ijkk} G_F}{8 \pi^2}
\left\{
\begin{array}{ll}
g^k_Q \frac{m_Z^2}{3} \ln \left( \frac{m_Z^2}{m_{NP}^2} \right) 
& {\rm massless} ~q_k \\ 
g_L^tm_t^2 \ln \left( \frac{m_t^2}{m_{NP}^2} \right) & q_k = t
\end{array}
\right. 
\eeq
where  $Q = L,R$ is the chirality of the quark
in the loop.
For light quarks,
this  gives $\delta g \sim  6 \times g^k_Q 10^{-3} \epsilon$.
Recall that  $g_L  = \pm  1 - 2 s_W^2 Q_{em}$, where $Q_{em}$
is the electric charge of the quark, and $g_R  = 2 s_W^2 Q_{em}$.
For a top quark in the loop, the $Z$ can always couple to 
the $t_L$ with $g_L \simeq 2/3$, because in the case where
$t_R$ participates in the four fermion vertex, the chirality
flip can be provided by $m_t$ without suppressing the loop.

To obtain the bounds in the tables, we require for $i = j$
\beq
\frac{2 g_Z^j \delta g_Z^{jj}}{(g_L^j)^2 + (g_R^j)^2} <
2 \frac{\delta R_{l_j}} {R_{l_j}} 
\eeq
(at ``two $\sigma$''), and
and for $i \neq j$
\beq
  \frac{1}{2.5 GeV} \frac{\sqrt{2} G_F m_Z^3}{6  \pi}
(\delta g^{ij} )^2 < BR(Z \to l_i \bar{l}_j) 
\eeq

\subsection{ Neglected Constraints}
The constraint arising from $\mu -e$ conversion \cite{Shanker:1979ap}  
is reproduced from \cite{Raidal:2008jk}, because  
the experimental value \cite{Dohmen:1993mp} has not changed.
Constraints can also be obtained from other processes,
such as loop contributions
to g-2 \cite{Cheung:2001ip} and  neutrino interactions
\cite{Doncheski:1997it,Grossman:1995wx,Honda:2007wv}.
We neglect bounds from oblique parameters \cite{precision},
because these depend on the leptoquark couplings to the Higgs,
which we do not consider. 

\section{The tables}
\label{sec:tables}

We quote bounds on  four fermion interactions of chiral fields.
Bounds on the coefficients of gauge invariant operators
an be extracted 
by expanding the operators as a sum of four fermion
interactions, then identifying from the tables the most restrictive
bound on the various four fermion interactions.
In the table \ref{tabs} are  listed the 
tables whose bounds apply to the coefficients of
the gauge invariant operators of eqns
\ref{O1lq} to \ref{4}.

\begin{table}
\begin{tabular}{l||c|c|c|c|c|c|c|c|c|c|c|c}
Operator & tab \ref{tab:eeddND}
 & tab \ref{tab:eeuuND}  & tab \ref{tab:ePLeuPLu}  & tab \ref{tab:eeuu}
  & tab \ref{tab:eeuPLu}  & tab \ref{tab:ePLeuu} & tab \ref{tab:ePLedPLd}
 & tab \ref{tab:eedd}  & tab \ref{tab:ePLedd}  & tab \ref{tab:eedPLd}
 & tab \ref{tab:nunuqq}  & tab \ref{tab:CC}\\
\hline\hline
${\cal O}_{(1) \ell q}$ & x &x & x & & & & x & & & & x & \\
${\cal O}_{(3) \ell q}$ & x &x & x & & & & x & & & & x & x\\
\hline
${\cal O}_{e q}$ & x &x &  & &x & &  & & &x &  & \\
${\cal O}_{e d}$ & x & &  & & & &  &x & & &  & \\
${\cal O}_{e u}$ &  &x &  &x & & &  & & & &  & \\
\hline
${\cal O}_{\ell u}$ &  &x &  & & &x &  & & & &x  & \\
${\cal O}_{\ell d}$ &x & &  & & & &  & &x & &x  & 
\end{tabular}
\caption{ Operators whose coefficients are constrained by
the bounds  in  tables \ref{tab:eeddND} to \ref{tab:CC}.}
\label{tabs}
\end{table}

Table  \ref{tab:eeddND}
 contains bounds which apply to 
$V\pm A$  generation-off-diagonal 
interactions involving charged leptons
and ups  of any chirality: 
$(\bar{l}^i \gamma^\mu P_{L,R} l^j)$ $
(\bar{u}^k \gamma^\mu P_{L,R} u^n)$. Table  \ref{tab:eeuuND} 
is the same for interactions involving $d$-type quarks:
 $(\bar{l}^i \gamma^\mu P_{L,R} l^j)
(\bar{d}^k \gamma^\mu P_{L,R} d^n)$. 
Notice that  $u$
and $d$  can be SU(2) singlets or doublets.
Table \ref{tab:SpmA} gives
bounds which apply to all the  $S \pm P $ operators. 
Many of the bounds in tables  \ref{tab:eeddND}, \ref{tab:eeuuND}
and \ref{tab:SpmA} arise from leptonic meson decays.
In tables \ref{tab:eeddND}, \ref{tab:eeuuND}
and \ref{tab:SpmA}, all the flavour off-diagonal index
combinations are listed. In the case where
the bound depends on the chirality of the operator
({\it e.g. } $Z \to l_i \bar{l}_j$), the
process is listed,  but the bounds are in the chirality-specific 
tables 
 \ref{tab:ePLeuPLu}, 
\ref{tab:eeuu},
\ref{tab:ePLeuu},
\ref{tab:eeuPLu},
\ref{tab:ePLedPLd}
\ref{tab:eedd},
\ref{tab:ePLedd},
\ref{tab:eedPLd},
and 
\ref{tab:nunuqq},
\ref{tab:CC}.

In the tables 
 \ref{tab:ePLeuPLu}, 
\ref{tab:eeuu},
\ref{tab:ePLeuu},
\ref{tab:eeuPLu},
\ref{tab:ePLedPLd}
\ref{tab:eedd},
\ref{tab:ePLedd},
\ref{tab:eedPLd}
and 
\ref{tab:CC},  are 
listed all the generation diagonal coefficients
$\epsilon^{iikk}$, and other chirality dependent
bounds, again on the coefficients of four fermion
interactions, rather than the coefficients of
gauge invariant operators.
 Tables  \ref{tab:ePLeuPLu} to  \ref{tab:eedPLd}
involve two charged leptons,
  whereas tables \ref{tab:nunuqq} and \ref{tab:CC}
constrain respectively interactions
containing  $(\bar{\nu} \gamma \nu)$, and charged current
interactions. The bounds on
four fermion interactions involving
  $(\bar{\nu} \gamma \nu)$ are incomplete
(see, {\it e.g.} \cite{Biggio} for a more
complete analysis), because we are interested
in dimension six contact interactions. 
  At dimension six, any gauge invariant operator
inducing  $(\bar{\nu} \gamma \nu)(\bar{f} \gamma f)$, where
$f \in \{ q,u,d \}$
also induces charged current,  or charged
lepton neutral current interactions,
whose coefficients are usually more strictly
constrained. So  table \ref{tab:nunuqq}
only contains a few  interesting constraints.

The first colomn of each table gives the generation
indices $ijkn$ of the operator.  The bounds 
are ordered from first to third generation in
each index, so they  
start with $ijkn$ = 1111,  then 1112 and so
on to 3333.  The second colomn  is
the numerical bound on $\epsilon^{ijkn}$, defined in eqn
(\ref{7}), arising from the observation listed in the
third colomn. In the last colomn, we give
the numerical value of the data we used; in the cases
where the bound can be expressed  $|\epsilon|^2 < BR^{expt} \times$
constant, then this allows  to rescale the bounds 
on $\epsilon$ when the data improves.

All the four fermion interactions except those of
table \ref{tab:CC}
are  hermitian, so  all bounds  apply
under simultaneous permutation of lepton and quark indices.
In many cases, the experimental bounds apply
under permutation on one pair of indices;
for instance, the bound from $D^+ \to \pi^+ e^\pm \mu^\pm$
 applies to $\epsilon^{e \mu cu}$  and  $\epsilon^{ \mu e cu}$. 
We assume that bounds on the decays 
$D^+ \to X$ and $B^+ \to X$ apply also to 
the CP conjuagate processes $D^- \to \overline{X}$ and $B^- \to \overline{X}$.

Several operators involving  a top quark 
are unconstrained, although they
would contribute to tree level top decay 
(Therefore  our tables are often missing the
rows corresponding to top quarks.).
One could hope for 
 a rough bound  $\epsilon < V_{tb}$ from  top
data   at the Tevatron.  However,
we did not find a Tevatron analysis 
of $t$ decay via a four-fermion operator,
where the kinematics  will
be different from the expected $t \to W b$.

\begin{table}[htp!]
\begin{center}
\begin{tabular}{|c|c|c|c|}
\hline
&&&\\
$\left(\bar{e}_{i}\gamma^{\mu}P_{L,R}e_{j}\right)\left(\bar{d}_{k}\gamma_{\mu} P_{L,R}d_{n}\right)$ 
 & Constraint on $\epsilon^{ijkn}$ & Observable &Experimental value\\
&&& \\
\hline
\hline
$eeds$   & $5.7\times10^{-5}$ & $BR(K^{0}_{L}\rightarrow \bar{e}e)$ & $9,0\times10^{-12}$\\
\hline
$eedb$   & $2.0\times10^{-4}$     & 
$\frac{BR(B^{+}\rightarrow \pi^{+}\bar{e}e)}
{{BR(B^{0}\rightarrow\pi^{-}\bar{e}\nu_{e})}}$ & 
$<\frac{1.8\times10^{-7}}{1.34\times10^{-4}}$\\
\hline
$eesb$   & $1.8\times10^{-4}$     
& $\frac{BR(B^{+}\rightarrow K^{+}\bar{e}e)}
{{BR(B^{+}\rightarrow D^{0}\bar{e}\nu_{e})}}$ 
& $\frac{4.9\times10^{-7}}{2.2\times10^{-2}}$\\
\hline
$e\mu dd$ & $8.5\times10^{-7}$ & $\mu-e$ conversion on Ti & $\frac{\sigma(\mu^{-}Ti\rightarrow e^{-}Ti)}{\sigma(\mu^{-}Ti\rightarrow capture)}< 4.3\times10^{-12}$\\
\hline
$e\mu ds$   & $3.0\times10^{-7}$   & $BR(K^{0}_{L}\rightarrow \bar{e}\mu)$ & $<4.7\times10^{-12}$\\
\hline
$e\mu db$ & $2.0\times10^{-4}$     
& $\frac{BR(B^{+}\rightarrow \pi^{+}\bar{e}\mu)}
{{BR(B^{0}\rightarrow\pi^{-}\bar{e}\nu_{e})}}$ 
& $<\frac{1.7\times10^{-7}}{1.34\times10^{-4}}$\\
\hline
$e\mu ss$ && $Z \to \bar{e} \mu$ &\\ 
\hline
$e\mu sb$ & $8 \times10^{-5}$       
& $\frac{BR(B^{+}\rightarrow K^{+}\bar{e}\mu)}
{{BR(B^{+}\rightarrow D^{0}\bar{e}\nu_{e})}}$ 
& $<\frac{9.1\times10^{-8}}{2.2\times10^{-2}}$\\
\hline
$e\mu bb$ && $Z \to \bar{e} \mu$ &\\ 
\hline
$e \tau  dd $ & $8.4\times10^{-4}$
 & $\frac{BR(\tau^{-}\rightarrow \pi^{0}e^{-})}
{BR(\tau^{-}\rightarrow \pi^{-}\nu_{\tau})}$ 
& $<\frac{8\times10^{-8}}{10.91\times10^{-2}}$\\
\hline
$e\tau ds$ & $4.9\times10^{-4} $    
& $\frac{BR( \tau \rightarrow {e} K)}
{BR( \tau \rightarrow \bar{\nu} K)}$ & $BR <3.3  \times10^{-8}$\\
\hline
$e\tau db$ & $4.1\times10^{-3}$    
& $BR(B^{0}\rightarrow \bar{e}\tau)$ 
& $<1.1\times10^{-4}$\\
\hline
$e\tau ss$ && $Z \to \bar{e} \tau$ &\\ 
\hline
$e\tau sb$ &     & $$ & \\
\hline
$e\tau bb$ && $Z \to \bar{e} \tau$ &\\ 
\hline
$\mu\mu ds$   & $7.8\times10^{-6}$ & $BR(K^{0}_{L}\rightarrow \bar{\mu}\mu)$ & $6.84\times10^{-9}$\\
\hline
$\mu\mu db$   & $1.3\times10^{-4}$     
& $\frac{BR(B^{+}\rightarrow \pi^{+}\bar{\mu}\mu)}
{{BR(B^{0}\rightarrow\pi^{-}\bar{e}\nu_{e})}}$ 
& $<\frac{6.9\times10^{-8}}{1.3\times10^{-4}}$\\
\hline
$\mu\mu sb$   & $7.0\times10^{-5}$     
& $\Gamma(B_s \rightarrow \bar{\mu}\mu)$ 
& $BR < 4.7\times10^{-8}$\\
\hline
$\mu\tau dd$ & $9.8\times10^{-4}$ & $\frac{BR(\tau^{-}\rightarrow \pi^{0}\mu^{-})}{BR(\tau^{-}\rightarrow \pi^{-}\nu_{\tau})}$ & $<\frac{1.1\times 10^{-7}}{10.91\times 10^{-2}}$\\
\hline
$\mu \tau ds$ & $5.4\times10^{-4} $    
& $\frac{BR( \tau \rightarrow {\mu} K)}
{BR( \tau \rightarrow \bar{\nu} K)}$ & $BR <4.0  \times10^{-8}$\\
\hline
$\mu\tau db$ & $2.1\times10^{-2}$  
& $BR(B^{0}\rightarrow \bar{\mu}\tau)$ 
& $<2.2\times10^{-5}$\\
\hline
$\mu \tau ss$ && $Z \to \bar{\mu} \tau$ &\\ 
\hline
$\mu \tau sb$ & $ 2.3\times10^{-3}$       
& $\frac{BR(B^+ \to K^+ \bar{\tau} \mu)}
{{BR(B^{+}\rightarrow D^{0}\bar{e}\nu)}}$ 
& $<\frac{7.7\times10^{-5}}{2.2\times10^{-2}}$\\
\hline
$\mu \tau bb$ && $Z \to \bar{\mu} \tau$ &\\ 
\hline
$\tau \tau ds$ &      & $$ &  \\
\hline
$\tau\tau db$   & $0.2 $  
& $BR(B^{0}\rightarrow \bar{\tau}\tau)$ 
& $<4.1\times10^{-3}$\\
\hline
$\tau \tau sb$ &        & $$  & \\
\hline

\end{tabular}
\caption{
Constraints on the dimensionless
coefficient  $\epsilon^{ijkn}$, of the four-fermion
interaction $2 \sqrt{2}  G_F \left(\bar{e}_{i} \gamma^{\mu} P_{L,R}
e_{j}\right)\left(\bar{d}_{k}\gamma_{\mu} P_{L,R} d_{n}\right) $. 
 $P_{L,R}$ can be $P_{L}$ or $P_{R}$. The
bounds  collected in this table are flavour-changing, and 
 apply to many   of the operators;
see the table \ref{tabs}.  The 
generation indices $ijkn$  are given in the first column, 
and the  best bound   in 
column two. It  arises from  the observable of column 3,
and  the experimental value we used is given  in column 4.
All bounds apply under permutation of the lepton and/or
quark indices.
}
\label{tab:eeddND}
\end{center}
\end{table}

\begin{table}[htp!]
\begin{center}
\begin{tabular}{|c|c|c|c|}
\hline
&&&\\
$\left(\bar{e}_{i} \gamma^{\mu} P_{L,R}
e_{j}\right)\left(\bar{u}_{k}\gamma_{\mu} P_{L,R} u_{n}\right)$
 & Constraint on $\epsilon^{ijkn}$ & Observable &Experimental value\\
&&& \\
\hline
\hline
$eeuc$   & $7.9\times10^{-3}$    & $\frac{BR(D^{+}\rightarrow \pi^{+}\bar{e}e)}{BR(D^{0}\rightarrow\pi^{-}\bar{e}\nu_{e})}$ & $<\frac{7.4\times10^{-6}}{2.83\times10^{-3}}$\\
\hline
$ee ut$ & $ $ &  & $$\\
\hline
$ee ct$ & $ $ &  & $$\\
\hline
$ee tt$ & .092 & $Z \to \bar{e} e$ &  $  R_e = 20.804
 \pm .050  $\\ 
\hline
$e\mu uu$ & $8.5\times10^{-7}$ & $\mu-e$ conversion on Ti & $\frac{\sigma(\mu^{-}Ti\rightarrow e^{-}Ti)}{\sigma(\mu^{-}Ti\rightarrow capture)}< 4.3\times10^{-12}$\\
\hline
$e\mu uc$ & $1.7\times10^{-2}$     & $\frac{BR(D^{+}\rightarrow \pi^{+}\bar{e}\mu)}{BR(D^{0}\rightarrow \pi^{-}\bar{e}\nu_{e})}$ & $<\frac{3.4\times10^{-5}}{2.83\times10^{-3}}$\\
\hline
$e\mu ut$ & & & $$\\
\hline
$e\mu cc$  &  &  $Z \to \bar{e} \mu$  & $$\\
%
\hline
$e\mu ct$ &  &  & $$\\
\hline
$e\mu tt$ & .1 & $Z \to \bar{e} \mu$ &  $BR<1.7 \times 10^{-6} $\\ 
\hline
$e \tau  uu$ & $8.4\times10^{-4}$ & $\frac{BR(\tau^{-}\rightarrow \pi^{0}e^{-})}{BR(\tau^{-}\rightarrow \pi^{-}\nu_{\tau})}$ & $<\frac{8\times10^{-8}}{10.91\times10^{-2}}$\\
\hline
$e \tau uc$ & &  & $$\\
\hline
$e \tau ut$ & &  & $$\\
\hline
$e \tau cc$  &  & $Z \to \bar{e} \tau$  & $$\\
\hline
$e \tau ct$ &  &  & $$\\
\hline
$e \tau tt$ & 
 0.2      &  $Z \to \bar{e} \tau$ & $BR< 9.8 \times 10^{-6}$\\
\hline
$\mu\mu uc$  & $6.1\times10^{-3}$    & $\frac{BR(D^{+}\rightarrow \pi^{+}\bar{\mu}\mu)}{BR(D^{0}\rightarrow \pi^{-}\bar{e}\nu_{e})}$ & $<\frac{3.9\times10^{-6}}{2.83\times10^{-3}}$\\
\hline
$ \mu \mu  ut$ & &  & $$\\
\hline
$ \mu \mu  ct$ & &  & $$\\
\hline
$\mu \mu  tt$ & .061 & $Z \to \bar{\mu} \mu$ &  $R_\mu =  20.785
 \pm .033  $\\ 
\hline
$\mu\tau uu$ & $9.8\times10^{-4}$ & $\frac{BR(\tau^{-}\rightarrow \pi^{0}\mu^{-})}{BR(\tau^{-}\rightarrow \pi^{-}\nu_{\tau})}$ & $<\frac{1.1\times10^{-7}}{10.91\times10^{-2}}$\\
\hline
$\mu \tau uc$ &  &  & $$\\
\hline
$\mu \tau ut$ &  &  & $$\\
\hline
$\mu \tau cc$  &  &$Z \to \tau \bar{\mu}$  & $$\\ 
\hline
$\mu \tau ct$ & &  & $$\\
\hline
$\mu \tau tt$ & 1 &$Z \to \tau \bar{\mu}$  & $BR< 12 \times 10^{-6}$\\
\hline
$\tau \tau uc$ &  &   & $$\\
\hline
$\tau \tau ut$ & &  & $$\\
\hline
$\tau \tau ct$ & &  & $$\\
\hline
$\tau \tau  tt$ & .086 & $Z \to \bar{\tau} \tau$ &  $R_\tau = 20.764 \pm.045 $\\ 
\hline
\end{tabular}
\caption{Flavour-changing constraints on the dimensionless
coefficient  $\epsilon^{ijkn}$, of the four-fermion
interaction $2 \sqrt{2}  G_F \left(\bar{e}_{i} \gamma^{\mu} P_{L,R}
e_{j}\right)\left(\bar{u}_{k}\gamma_{\mu} P_{L,R} u_{n}\right) $. 
 $P_{L,R}$ can be $P_{L}$ or $P_{R}$.  
 The
bounds  collected in this table  apply to many  of the operators;
see the table \ref{tabs}.
The 
generation indices $ijkn$  are given in the first column, 
and the  best bound   in 
column two. It  arises from  the observable of column 3,
and  the experimental value we used is given  in column 4.
All bounds  apply under  permutation of lepton and quark indices.
}
\label{tab:eeuuND}
\end{center}
\end{table}

\begin{table}[htp!]
\begin{center}
\begin{tabular}{|c|c|c|c|}
\hline
&&&\\
$\left(\bar{e}_{i}\gamma^{\mu}P_L e_{j}\right)\left(\bar{u}_{k}\gamma_{\mu}P_L u_{n}\right)$  & Constraint on $\epsilon^{ijkn}_{(n) \ell q}$ & Observable &Experimental value\\
&&& \\
\hline
\hline

$eeuu$ & $> 1 \times 10^{-2}$  & $\Lambda^{\pm}_{eeuu;LL}$ &  see sec \ref{contint} \\
\hline
%
$eecc$ & $< 1 \times10^{-2}$  & $\Lambda^{\pm}_{eecc;LL}$ &$ >4.4,5.6$ ~ TeV\\
\hline
$ee tt $ & $0.092 $  &  $Z\to e \bar{e}  $  & $R_e =  20.804
 \pm .050$ \\
\hline
$e\mu cc$ & $ .6$       & $Z \to \bar{e} \mu$ & $BR<1.7 \times 10^{-6} $\\
\hline
$e \tau cc$ & $ 1.4$    &  $Z \to \bar{\mu} \tau$ &  $BR< 9.8 \times 10^{-6}$\\
\hline

$\mu \mu uu $&
 $> -2.0\times 10^{-2}$  & $\Lambda^{+}_{\mu \mu qq;LL}$ & $>2.9$ ~ TeV\\
& $< 7.8\times 10^{-3}$  & $\Lambda^{-}_{\mu \mu qq;LL}$ & $>4.2$ ~ TeV\\
\hline
$\mu \mu cc $ & $0.4  $  &  $Z \to \mu \bar{\mu} $  & $R_\mu < 20.785 \pm .033$  \\
\hline

$\mu \mu t t $ & $0.061  $  &  $Z \to \mu \bar{\mu} $  & $R_\mu <  20.785 \pm .033$  \\
\hline
$\mu \tau cc$ & $ 1.6$    &  $Z \to \bar{\mu} \tau$ &  $BR< 12 \times 10^{-6}$\\
\hline
%
%
$\tau \tau uu $ & $0.54  $  &  $Z \to \tau \bar{\tau} $  & $R_\tau < 20.765 \pm .045$  \\
\hline
$\tau \tau cc $ & $0.54  $  &  $Z \to \tau \bar{\tau} $  & $R_\tau < 20.765 \pm .045$  \\
\hline

$\tau \tau t t $ & $0.086  $  &  $Z \to \tau \bar{\tau} $  & $R_\tau < 20.765 \pm .045$  \\
\hline
\end{tabular}
\caption{
Flavour diagonal constraints on the dimensionless
coefficient  $\epsilon^{ijkn}$, of the four-fermion
interaction $2 \sqrt{2}  G_F \left(\bar{e}_{i} \gamma^{\mu} P_{L}
e_{j}\right)\left(\bar{u}_{k}\gamma_{\mu} P_{L} u_{n}\right) $.
 The
 indices $ijkn$ are  given  in the left colomn,
the  best calculated constraints  are in
column 2, arising the  observable  of column 3, with the  
experimental value in column 4. 
All bounds  apply
under  permutation of lepton and/or quark indices.
}
\label{tab:ePLeuPLu}
\end{center}
\end{table}

\begin{table}[!tpb]
\begin{center}
\begin{tabular}{|c|c|c|c|}
\hline
&&&\\
$\left(\bar{e}_{i}\gamma^{\mu}P_{R}e_{j}\right)\left(\bar{u}_{k}\gamma_{\mu}P_{R}u_{n}\right)$ & Constraint on $\epsilon^{ijkn}_{eu}$ & Observable & Experimental value\\
&&& \\
\hline
\hline
$eeuu$ & $1\times10^{-2}$  & $\Lambda^{\pm}_{eeuu;RR}$ & see sec \ref{contint} \\
\hline
$eecc$ & $>-9 \times10^{-2}$  & $\Lambda^{+}_{eecc;RR}$ &$ >1.5$ ~ TeV \\
 & $<1.8\times10^{-2}$  & $\Lambda^{-}_{eecc;RR}$ &$ >3.8$ ~ TeV \\
\hline
$ee t t $ & $0.092 $  & $Z \to e \bar{e} $  & $R_e < 20.804 \pm .050$  \\
\hline
$e\mu cc$ & $ 1.2$       & $Z \to \bar{e} \mu$ & $BR<1.7 \times 10^{-6} $\\
\hline
%
%
$e\tau cc$ & 2.8  &  $Z \to \bar{e} \tau$ & $BR< 9.8 \times 10^{-6}$\\
\hline
%
%
$\mu\mu uu$ & $>-2.2\times10^{-2}$  & $\Lambda^{+}_{\mu\mu uu;RR}$ & $>4.2$ ~ TeV\\
 & $<8.5 \times10^{-3}$  & $\Lambda^{-}_{\mu\mu uu;RR}$ & $>6.7$ ~ TeV\\
\hline
$\mu \mu cc $ & $0.79  $  & $Z \to \mu \bar{\mu} $  & $R_\mu < 20.785 \pm .033$  \\
\hline
$\mu \mu t t $ & $0.061  $  & $Z \to \mu \bar{\mu} $  & $R_\mu < 20.785 \pm .033$  \\
\hline
$\mu \tau cc$ & $  3$    &  $Z \to \bar{\mu} \tau$ &  $BR< 12 \times 10^{-6}$\\
\hline
%
%
$\tau \tau uu $ & $1.1  $  & $Z \to \tau \bar{\tau} $  & $R_\tau = 20.764 \pm.045$  \\
\hline
$\tau \tau cc $ & $1.1$  & $Z \to \tau \bar{\tau} $  & $R_\tau = 20.764 \pm.045$  \\
\hline
$\tau \tau t t $ & $0.086  $  &$Z \to \tau \bar{\tau} $  & $R_\tau = 20.764 \pm.045$  \\

\hline
\end{tabular}
\end{center}
\caption{
Flavour diagonal constraints on the dimensionless
coefficient  $\epsilon^{ijkn}_{eu}$, of the four-fermion
interaction $2 \sqrt{2}  G_F \left(\bar{e}_{i} \gamma^{\mu} P_{R}
e_{j}\right)\left(\bar{u}_{k}\gamma_{\mu} P_{R} u_{n}\right) $.
 The
 indices $ijkn$ are  given  in the left colomn,
the  best calculated constraints  are in
column 2, arising the  observable  of column 3, with the  
experimental value in column 4. 
All bounds  apply
under  permutation of lepton and/or quark indices.
}
\label{tab:eeuu}
\end{table}

\begin{table}[!tpb]
\begin{center}
\begin{tabular}{|c|c|c|c|}
\hline
&&&\\
$\left(\bar{e}_{i}\gamma^{\mu}P_{R}e_{j}\right)\left(\bar{u}_{k}\gamma_{\mu}P_{L}u_{n}\right)$ & Constraint on $\epsilon^{ijkn}_{eq}$ & Observable & Experimental value\\
&&& \\
\hline
\hline
$eeuu$ & $1.2\times10^{-2}$  & $\Lambda^{\pm}_{eeuu;LR}$ & see sec \ref{contint} \\
\hline
$eecc$ & $ 4.2 \times10^{-2}$  & $\Lambda^{\pm}_{eecc;LR}$ &$ \gsim 3$ ~ TeV \\
\hline
$ee t t $ & $0.092 $  & $Z \to e \bar{e} $  & $R_\tau < 20 \pm .08$  \\
\hline
$e\mu cc$ & $ 0.6$       & $Z \to \bar{e} \mu$ & $BR<1.7 \times 10^{-6} $\\
\hline
%
%
$e\tau cc$ & 1.4 &  $Z \to \bar{e} \tau$ & $BR< 9.8 \times 10^{-6}$\\
\hline
%
%
$\mu\mu uu$ & $<1.5\times10^{-2}$  & $\Lambda^{\pm}_{\mu\mu uu;RL}$ & $\gsim 5.2$ ~ TeV\\
\hline
$\mu \mu cc $ & $0.40  $  & $Z \to \mu \bar{\mu} $  & $R_\mu < 20.785 \pm .033$  \\
\hline
$\mu \mu t t $ & $0.061  $  & $Z \to \mu \bar{\mu} $  & $R_\mu < 20.785 \pm .033$  \\
\hline
$\mu \tau cc$ & $ 1.6$    &  $Z \to \bar{\mu} \tau$ &  $BR< 12 \times 10^{-6}$\\
\hline
%
%
$\tau \tau uu $ & $0.54  $  & $Z \to \tau \bar{\tau} $  & $R_\tau = 20.764 \pm.045$  \\
\hline
$\tau \tau cc $ & $0.54  $  & $Z \to \tau \bar{\tau} $  & $R_\tau = 20.764 \pm.045$  \\
\hline
$\tau \tau t t $ & $0.086  $  &$Z \to \tau \bar{\tau} $  & $R_\tau = 20.764 \pm.045$  \\

\hline
\end{tabular}
\end{center}
\caption{
Flavour diagonal constraints on the dimensionless
coefficient  $\epsilon^{ijkn}$, of the four-fermion
interaction $2 \sqrt{2}  G_F \left(\bar{e}_{i} \gamma^{\mu} P_{R}
e_{j}\right)\left(\bar{u}_{k}\gamma_{\mu} P_{L} u_{n}\right) $.
 The
 indices $ijkn$ are  given  in the left colomn,
the  best calculated constraints  are in
column 2, arising the  observable  of column 3, with the  
experimental value in column 4. 
 All bounds  apply
under  permutation of lepton and/or quark indices.
}
\label{tab:eeuPLu}
\end{table}
%

\begin{table}[!tpb]
\begin{center}
\begin{tabular}{|c|c|c|c|}
\hline
&&&\\
$  \left(\bar{e}_{i}\gamma^{\mu}P_L e_{j}\right)\left(\bar{u}_{k}\gamma_{\mu} P_R u_{n}\right)$ & Constraint on $\epsilon^{ijkn}_{\ell u}$ & Observable & Experimental value\\
&&& \\
\hline
\hline
$eeuu$ & $1.4\times10^{-2}$  & $\Lambda^{\pm}_{eeuu;LR}$ & see sec \ref{contint} \\
\hline
$eecc$ & $>-8.6 \times10^{-2}$  & $\Lambda^{+}_{eecc;LR}$ & $>2.1$ ~ TeV\\
 & $<2.6\times10^{-2}$  & $\Lambda^{-}_{eecc;LR}$ & $>3.4$ ~ TeV\\
\hline
$ee t t $ & $0.092 $  & $Z \to e \bar{e} $  & $R_e <  20.804
 \pm .050$  \\
\hline
$e\mu cc$ & $ 2.4$       & $Z \to \bar{e} \mu$ & $BR<1.7 \times 10^{-6} $\\
\hline
%
%
$e\tau cc$ &  2.8  &  $Z \to \bar{e} \tau$ & $BR< 9.8 \times 10^{-6}$\\
\hline
%
%
$\mu\mu uu$ & $1.4\times10^{-2}$  & $\Lambda^{\pm}_{\mu\mu uu;LR}$ &$ >5.2$ ~ TeV\\
\hline
$\mu \mu cc $ & $0.79  $  & $Z \to \mu \bar{\mu} $  & $R_\mu < 20.785 \pm .033$  \\
\hline
$\mu \mu t t $ & $0.061  $  & $Z \to \mu \bar{\mu} $  & $R_\mu < 20.785 \pm .033$  \\
\hline
$\mu \tau cc$ & $ 3$    &  $Z \to \bar{\mu} \tau$ &  $BR< 12 \times 10^{-6}$\\
\hline
%
%
$\tau \tau uu $ & $ 1.1 $  & $Z \to \tau \bar{\tau} $  & $R_\tau = 20.764 \pm.045$  \\
\hline
$\tau \tau cc $ & $1.1 $  & $Z \to \tau \bar{\tau} $  & $R_\tau < 20.764 \pm .045$  \\
\hline
$\tau \tau t t $ & $0.086  $  &$Z \to \tau \bar{\tau} $  & $R_\tau = 20.764 \pm.045$  \\
\hline

\end{tabular}
\caption{
Flavour diagonal constraints on the dimensionless
coefficient  $\epsilon^{ijkn}$, of the four-fermion
interaction $2 \sqrt{2}  G_F \left(\bar{e}_{i} \gamma^{\mu} P_{L}
e_{j}\right)\left(\bar{u}_{k}\gamma_{\mu} P_{R} u_{n}\right) $.
 The
 indices $ijkn$ are  given  in the left colomn,
the  best calculated constraints  are in
column 2, arising the  observable  of column 3, with the  
experimental value in column 4. 
All bounds  apply
under  permutation of lepton and/or quark indices.
 }
\label{tab:ePLeuu}
\end{center} 
\end{table}

\begin{table}[htp!]
\begin{center}
\begin{tabular}{|c|c|c|c|}
\hline
&&&\\
$\left(\bar{e}_{i}\gamma^{\mu}P_L e_{j}\right)
\left(\bar{d}_{k}\gamma_{\mu}P_L d_{n}\right)$  
& Constraint on $\epsilon^{ijkn}_{(n) \ell q }$ & Observable &Experimental value\\
&&& \\
\hline
\hline

$eedd$ & $ 1.3 \times 10^{-2}$  & $\Lambda^{\pm}_{eedd;LL}$ &  see sec \ref{contint} \\
\hline
%
$eess$ & $ 1.5 \times10^{-2}$  & $\Lambda^{\pm}_{eeqq;LL}$ &$ >9.7, 8.0$ ~ TeV\\
 &  &  &  see sec \ref{contint} \\
\hline
$ee bb $ &  $ >-4.3 \times10^{-3}$  & $\Lambda^{+}_{eebb;LL}$ &$ >9.4$ ~ TeV\\
 &  $ <1.5 \times10^{-2}$  & $\Lambda^{-}_{eebb;LL}$ &$ >4.9$ ~ TeV\\
\hline
$e\mu ss$ & $ .5$       & $Z \to \bar{e} \mu$ & $BR<1.7 \times 10^{-6} $\\
\hline
$e\mu bb$ & $ .5$       & $Z \to \bar{e} \mu$ & $BR<1.7 \times 10^{-6} $\\
\hline
$e \tau ss$ & $ 1$    &  $Z \to \bar{e} \tau$ &  $BR< 9.8 \times 10^{-6}$\\
\hline
$e \tau bb$ & $ 1$    &  $Z \to \bar{e} \tau$ &  $BR< 9.8 \times 10^{-6}$\\
\hline
$\mu \mu dd$&
 $> -6.4\times 10^{-2}$  & $\Lambda^{+}_{\mu \mu dd;LL}$ & $>4.2$ TeV \\
& $< 2.3 \times 10^{-2}$  & $\Lambda^{-}_{\mu \mu dd;LL}$ &$ > 7.0$ TeV \\
\hline
$\mu \mu ss $ & $0.32  $  &  $Z \to \mu \bar{\mu} $  & $R_\mu < 20.785 \pm .033$  \\
\hline

$\mu \mu bb $ & $0.32  $  &  $Z \to \mu \bar{\mu} $  & $R_\mu < 20.785 \pm .033$  \\
\hline
$\mu \tau ss$ & $ 1$    &  $Z \to \bar{\mu} \tau$ &  $BR< 12 \times 10^{-6}$\\
\hline
$\mu \tau bb$ & $ 1$    &  $Z \to \bar{\mu} \tau$ &  $BR< 12 \times 10^{-6}$\\
\hline
%
%
$\tau \tau dd $ & $ .43$  &  $Z \to \tau \bar{\tau} $  & $R_\tau = 20.764 \pm.045$  \\
\hline
$\tau \tau ss $ & $.43  $  &  $Z \to \tau \bar{\tau} $  & $R_\tau = 20.764 \pm.045$  \\
\hline

$\tau \tau bb $ & $.43  $  &  $Z \to \tau \bar{\tau} $  & $R_\tau = 20.764 \pm.045$  \\
\hline
\end{tabular}
\caption{
Flavour diagonal constraints on the dimensionless
coefficient  $\epsilon^{ijkn}$, of the four-fermion
interaction $2 \sqrt{2}  G_F \left(\bar{e}_{i} \gamma^{\mu} P_{L}
e_{j}\right)\left(\bar{d}_{k}\gamma_{\mu} P_{L} d_{n}\right) $.
 The
 indices $ijkn$ are  given  in the left colomn,
the  best calculated constraints  are in
column 2, arising the  observable  of column 3, with the  
experimental value in column 4. 
 All bounds  apply
under  permutation of lepton and/or quark indices.
}
\label{tab:ePLedPLd}
\end{center}
\end{table}

\begin{table}[!tpb]
\begin{center}
\begin{tabular}{|c|c|c|c|}
\hline
&&&\\
$\left(\bar{e}_{i}\gamma^{\mu}P_{R}e_{j}\right)\left(\bar{d}_{k}\gamma_{\mu}P_{R}d_{n}\right)$ & Constraint on $\epsilon^{ijkn}_{ed}$ & Observable & Experimental value\\
&&& \\
\hline
\hline
$eedd$ & $2.2\times10^{-2}$  & $\Lambda^{\pm}_{eedd;RR}$ &  see sec \ref{contint}\\
\hline
$ee ss  $ &  $>-2.0\times10^{-2}$  & $\Lambda^{+}_{eeqq;RR}$ & $7.6$ TeV \\
$ee ss  $ &  $<3.7\times10^{-2}$  & $\Lambda^{-}_{eeqq;RR}$ & $5.6$ TeV\\
\hline
$eebb$ & $>-1.5\times10^{-2}$  & $\Lambda^{+}_{eebb;RR}$ & $4.9$~ TeV\\
 & $5.6\times10^{-2}$  & $\Lambda^{-}_{eebb;RR}$ & $2.6$~ TeV\\
\hline
$e\mu ss$ & $ 2.4$       & $Z \to \bar{e} \mu$ & $BR<1.7 \times 10^{-6} $\\
\hline
$e\mu bb$ & $ 2.4$       &  $Z \to \bar{e} \mu$ &  $BR<1.7 \times 10^{-6} $\\
\hline
$e\tau ss$ & 6      &  $Z \to \bar{e} \tau$ & $BR< 9.8 \times 10^{-6}$\\
\hline
$e\tau bb$ & $ 6 $    &  $Z \to \bar{e} \tau$ &   $BR< 9.8 \times 10^{-6}$\\
\hline
$\mu\mu dd$ & $>-1.7 \times10^{-1}$  & $\Lambda^{+}_{\mu\mu dd;RR}$ & $4.2$~ TeV\\
 & $<6.7\times10^{-2}$  & $\Lambda^{-}_{\mu\mu dd;RR}$ & $6.7$~ TeV\\
\hline
$\mu \mu ss  $ & $1.6  $  & $Z \to \mu \bar{\mu} $  & $R_\mu < 20.785 \pm .033$  \\
\hline
$\mu \mu bb  $ & $1.6 $  & $Z \to \mu \bar{\mu} $  & $R_\mu < 20.785 \pm .033$  \\
\hline
$\mu \tau ss$ & $ 6$    &  $Z \to \bar{\mu} \tau$ &  $BR< 12 \times 10^{-6}$\\
\hline
$\mu \tau bb$ & $ 6$    &  $Z \to \bar{\mu} \tau$ &  $BR< 12 \times 10^{-6}$\\
\hline
$\tau\tau dd  $ & $2.2  $  & $Z \to \tau \bar{\tau} $  & $R_\tau = 20.764 \pm.045$  \\
\hline
$\tau\tau ss  $ & $2.2  $  & $Z \to \tau \bar{\tau} $  & $R_\tau = 20.764 \pm.045$  \\
\hline
$\tau\tau bb  $ & $2.2 $  & $Z \to \tau \bar{\tau} $  & $R_\tau = 20.764 \pm.045$  \\

\hline
\end{tabular}
\end{center}
\caption{Flavour diagonal constraints on 
 on the dimensionless
coefficient  $\epsilon_{ed}^{ijkn}$, of the four-fermion
interaction $2 \sqrt{2}  G_F \left(\bar{e}_{i} \gamma^{\mu} P_{L}
e_{j}\right)\left(\bar{d}_{k}\gamma_{\mu} P_{L} d_{n}\right) $.
 The
 indices $ijkn$ are  given  in the left colomn,
the  best calculated constraints  are in
column 2, arising the  observable  of column 3, with the  
experimental value in column 4. 
 All bounds  apply
under  permutation of lepton and/or quark indices.
}
\label{tab:eedd} 
\end{table}

\begin{table}[!tpb]
\begin{center}
\begin{tabular}{|c|c|c|c|}
\hline
&&&\\
$  \left(\bar{e}_{i}\gamma^{\mu}P_L e_{j}\right)\left(\bar{d}_{k}\gamma_{\mu}P_Rd_{n}\right)$ & Constraint on $\epsilon^{ijkn}_{\ell d}$ & Observable & Experimental value\\
&&& \\
\hline
\hline
$eedd$ & $2.7\times10^{-2}$  & $\Lambda^{\pm}_{eedd;LR}$ & see sec \ref{contint}\\
\hline
$ee ss  $& $>-6.8\times10^{-2}$  & $\Lambda^{+}_{eeqq;LR}$ & 4.1 TeV\\
& $< 4.3\times10^{-2}$  & $\Lambda^{-}_{eeqq;LR}$ & 5.2 TeV\\
\hline
$eebb$ & $>-2.5\times10^{-2}$  & $\Lambda^{+}_{eebb;LR}$ &$ 3.9 $~ TeV\\
& $<5.0\times10^{-2}$  & $\Lambda^{-}_{eebb;LR}$ &$ 2.8 $~ TeV\\
\hline
$e\mu ss$ & $ 2.4$       & $Z \to \bar{e} \mu$ & $BR<1.7 \times 10^{-6} $\\
\hline
$e\mu bb$ & $ 2.4$       &  $Z \to \bar{e} \mu$ &  $BR<1.7 \times 10^{-6} $\\
\hline
$e\tau ss$ & 6      &  $Z \to \bar{e} \tau$ & $BR< 9.8 \times 10^{-6}$\\
\hline
$e\tau bb$ & $ 6 $    &  $Z \to \bar{e} \tau$ &   $BR< 9.8 \times 10^{-6}$\\
\hline
$\mu\mu dd$ & $1.2\times10^{-1}$  & $\Lambda^{\pm}_{\mu\mu dd;LR}$ & $5.2 $~ TeV\\
\hline
$\mu \mu ss $ & $1.6  $  & $Z \to \mu \bar{\mu} $  & $R_\mu < 20.785 \pm .033$  \\
\hline
$\mu \mu bb $ & $1.6  $  & $Z \to \mu \bar{\mu} $  & $R_\mu < 20.785 \pm .033$  \\
\hline
$\mu \tau ss$ & $ 6$    &  $Z \to \bar{\mu} \tau$ &  $BR< 12 \times 10^{-6}$\\
\hline
$\mu \tau bb$ & $ 6$    &  $Z \to \bar{\mu} \tau$ &  $BR< 12 \times 10^{-6}$\\
\hline
$\tau \tau dd $ & $2.2  $  & $Z \to \tau \bar{\tau} $  & $R_\tau = 20.764 \pm.045$  \\
\hline
$\tau \tau ss $ & $2.2  $  & $Z \to \tau \bar{\tau} $  & $R_\tau = 20.764 \pm.045$  \\
\hline
$\tau \tau bb $ & $2.2  $  & $Z \to \tau \bar{\tau} $  & $R_\tau = 20.764 \pm.045$  \\
\hline
\end{tabular}
\caption{
Flavour diagonal constraints on the dimensionless
coefficient  $\epsilon^{ijkn}$, of the four-fermion
interaction $2 \sqrt{2}  G_F \left(\bar{e}_{i} \gamma^{\mu} P_{L}
e_{j}\right)\left(\bar{d}_{k}\gamma_{\mu} P_{R} d_{n}\right) $.
 The
 indices $ijkn$ are  given  in the left colomn,
the  best calculated constraints  are in
column 2, arising the  observable  of column 3, with the  
experimental value in column 4. 
All bounds  apply
  under lepton and/or  quark index permutation.
}
\label{tab:ePLedd}
\end{center}
\end{table}

\begin{table}[!tpb]
\begin{center}
\begin{tabular}{|c|c|c|c|}
\hline
&&&\\
$  \left(\bar{e}_{i}\gamma^{\mu}P_R e_{j}\right)\left(\bar{d}_{k}\gamma_{\mu}P_L d_{n}\right)$ & Constraint on $\epsilon^{ijkn}_{eq}$ & Observable & Experimental value\\
&&& \\
\hline
\hline
$eedd$ & $2.2\times10^{-2}$  & $\Lambda^{\pm}_{eedd;RL}$ & see sec \ref{contint}\\
\hline
$ee ss  $& $>-8.0\times10^{-2}$  & $\Lambda^{+}_{eeqq;RL}$ & 3.8 TeV\\
& $< 3.2\times10^{-2}$  & $\Lambda^{-}_{eeqq;RL}$ & 6.0 TeV\\
\hline
$eebb$ & $>-6.6\times10^{-2}$  & $\Lambda^{+}_{eebb;RL}$ &$ 2.4 $~ TeV\\
& $<1.8\times10^{-2}$  & $\Lambda^{-}_{eebb;RL}$ &$ 4.6 $~ TeV\\
\hline
$e\mu ss$ & $ 0.5$       & $Z \to \bar{e} \mu$ & $BR<1.7 \times 10^{-6} $\\
\hline
$e\mu bb$ & $ 0.5$       &  $Z \to \bar{e} \mu$ &  $BR<1.7 \times 10^{-6} $\\
\hline
$e\tau ss$ & 1    &  $Z \to \bar{e} \tau$ & $BR< 9.8 \times 10^{-6}$\\
\hline
$e\tau bb$ & $ 1 $    &  $Z \to \bar{e} \tau$ &   $BR< 9.8 \times 10^{-6}$\\
\hline
$\mu\mu dd$ & $1.2\times10^{-1}$  & $\Lambda^{\pm}_{\mu\mu dd;LR}$ & $5.2 $~ TeV\\
\hline
$\mu \mu ss $ & $ 0.32  $  & $Z \to \mu \bar{\mu} $  & $R_\mu < 20.785 \pm .033$  \\
\hline
$\mu \mu bb $ & $ 0.32$  & $Z \to \mu \bar{\mu} $  & $R_\mu < 20.785 \pm .033$  \\
\hline
$\mu \tau ss$ & $ 1$    &  $Z \to \bar{\mu} \tau$ &  $BR< 12 \times 10^{-6}$\\
\hline
$\mu \tau bb$ & $ 1.8$    &  $Z \to \bar{\mu} \tau$ &  $BR< 12 \times 10^{-6}$\\
\hline
$\tau \tau dd $ & $0.43  $  & $Z \to \tau \bar{\tau} $  & $R_\tau = 20.764 \pm.045$  \\
\hline
$\tau \tau ss $ & $0.43  $  & $Z \to \tau \bar{\tau} $  & $R_\tau = 20.764 \pm.045$  \\
\hline
$\tau \tau bb $ & $0.43 $  & $Z \to \tau \bar{\tau} $  & $R_\tau = 20.764 \pm.045$  \\
\hline
\end{tabular}
\caption{
Flavour diagonal constraints on the dimensionless
coefficient  $\epsilon^{ijkn}$, of the four-fermion
interaction $2 \sqrt{2}  G_F \left(\bar{e}_{i} \gamma^{\mu} P_{R}
e_{j}\right)\left(\bar{d}_{k}\gamma_{\mu} P_{L} d_{n}\right) $.
 The
 indices $ijkn$ are  given  in the left colomn,
the  best calculated constraints  are in
column 2, arising the  observable  of column 3, with the  
experimental value in column 4. 
 All bounds  apply
   under lepton and/or  quark index permutation.
}
\label{tab:eedPLd}
\end{center}
\end{table}

\begin{table}[htp!]
\begin{center}
\begin{tabular}{|c|c|c|c|}
\hline
&&&\\
$\left(\bar{\nu}_{i}\gamma^{\mu}P_L \nu_{j}\right)\left(\bar{q}_{k}\gamma_{\mu}P_Lq_{n}\right)$  & Constraint on $\epsilon^{ijkn}$ & Observable &Experimental value\\
&&& \\
\hline
\hline

$\nu_\mu\nu_\mu q_1q_1$ & $7.3\times10^{-3}$  & $\Lambda^{-}_{\nu\nu qq;LL}$ &
$ >5.1$~ TeV \\

\hline

$\nu_i\nu_j ds$ & $9.4\times10^{-6}$     
& $\frac{BR(K^{+}\rightarrow \pi^{+}\bar{\nu}\nu)}
{BR(K^{+}\rightarrow\pi^{0}\bar{e}\nu_{e})}$ 
& $\frac{1.5\times10^{-10}}{5.08\times10^{-2}}$\\

\hline

$\nu_i\nu_j bd$ & $4.9\times10^{-2}$     
& $\frac{BR(B^{+}\rightarrow \pi^{+}\bar{\nu}\nu)}
{BR(B^{0}\rightarrow\pi^{+}\bar{e}\nu_{e})}$ 
& $<\frac{1.0\times10^{-4}}{1.34\times10^{-4}}$\\

\hline

$\nu_i\nu_j bs$ & $1.0 \times10^{-3}$     
& $\frac{BR(B^{+}\rightarrow K^{+}\bar{\nu}\nu)}
{BR(B^{+}\rightarrow D^{0}\bar{e}\nu_{e})}$ 
& $<\frac{1.4\times10^{-5}}{2.2\times10^{-2}}$\\
\hline

\end{tabular}
\caption{
Some constraints on the dimensionless
coefficient  $\epsilon^{ijkn}$, of the four-fermion
interaction $2 \sqrt{2}  G_F \left(\bar{\nu}_{i} \gamma^{\mu} P_{L}
\nu_{j}\right)\left(\bar{q}_{k}\gamma_{\mu} P_{L,R} q_{n}\right) $.
The indices $ijkn$ are  given  in the left colomn,
where is is specified whether $q$ can be  a $u$-type or $d$-type
quark. $q_1$ is a $u$ or  $d$. 
See \cite{Biggio} for a more complete list of constraints;
here are listed only those which can give
more restrictive  constraints  than
charged lepton interactions on the coefficients
of gauge-invariant dimension six operators.
}
\label{tab:nunuqq}
\end{center}
\end{table}

\begin{table}[htp!]
\begin{center}
\begin{tabular}{|c|c|c|c|}
\hline
&&&\\
$\left(\bar{\nu}_{i}\gamma^{\mu}P_L e_{j}\right)\left(\bar{d}_{k}\gamma_{\mu}P_Lu_{n}\right)$  & Constraint on $\epsilon^{ijkn}$ & Observable &Experimental value\\
&&& \\
\hline
\hline

$\nu_{e}edu$  & $1.0 \times10^{-3}$   & unitarity& $\delta V_{us} = .0019$\\
\hline
$\nu_{i}edu$  & $4.5 \times10^{-2}$   & unitarity& $\delta V_{us} = .0019$\\
\hline
$\nu_{e}edc$  &   $4.8 \times10^{-2}$  &unitarity&$\delta V_{cd} = .024$\\
\hline
$\nu_{i}edc$  &   $1.5 \times10^{-1}$ &unitarity&$\delta V_{cd} = .024$\\ 
\hline
$\nu_{e}e su$   &   $4.0 \times10^{-3}$  & unitarity& $\delta V_{us} = .0019$\\
\hline
$\nu_{i}e su$  &   $3.0 \times10^{-2}$  & unitarity& $\delta V_{us} = .0019$\\
\hline
%
$\nu_{e}esc$   & $1.2 \times10^{-1}$   & unitarity & $\delta V_{cs} = .06$\\
\hline
%
$\nu_{i} e sc$  & $2.4 \times10^{-1}$   & unitarity & $\delta V_{cs} = .06$\\
\hline
$\nu_{i}ebu$   & $3.9 \times 10^{-3}$ & $ \lsim V_{ub}$ & $<3.93\times10^{-3}$\\
\hline
$\nu_{i}ebc$  & $4.1 \times 10^{-2}$ & $ \lsim V_{cb}$ & $<4.12\times10^{-2}$\\
\hline

$\nu_{\mu}\mu du$   & $4.0\times10^{-3}$ & $R_{\pi}$ & $(1.230\pm 0.004)\times10^{-4}$\\
\hline

$\nu_{i}\mu du$   &$9.0 \times10^{-2}$ & $R_{\pi}$& $(1.230\pm 0.004)\times10^{-4}$\\
\hline
$\nu_{\mu}\mu dc$  &   $2.2 \times10^{-2}$  &unitarity&$\delta V_{cd} = .011$\\
\hline
$\nu_{i}\mu dc$  &   $1.0 \times10^{-1}$ &unitarity&$\delta V_{cd} = .011$\\ 
\hline
$\nu_{\mu}\mu su$  &   $4.0 \times10^{-3}$  & unitarity& $\delta V_{us} = .002$\\
\hline
$\nu_{i}\mu su$  &   $3.0 \times10^{-2}$  & unitarity& $\delta V_{us} = .002$\\

\hline

$\nu_{\mu}\mu sc$   & $1.2 \times10^{-1}$   & unitarity & $\delta V_{cs} = .06$\\
\hline
%
$\nu_{i} \mu sc$  & $2.4 \times10^{-1}$   & unitarity & $\delta V_{cs} = .06$\\
\hline
$\nu_{i} \mu bu$   & $3.9 \times 10^{-3}$ & $ \lsim V_{ub}$ & $<3.93\times10^{-3}$\\
\hline
$\nu_{i} \mu bc$  & $4.1 \times 10^{-2}$ & $ \lsim V_{cb}$ & $<4.12\times10^{-2}$\\
\hline
$\nu_{\tau} \tau du $  &  $1.0\times10^{-2}$     &$g_\tau/g_\mu \cite{Pich} $  
& $ 0.996 \pm .005 $\\
\hline
$\nu_{i} \tau du $  &  $2.0\times10^{-1}$    &$g_\tau/g_\mu \cite{Pich} $  
& $ 0.996 \pm .005 $\\
\hline
$\nu_{i} \tau dc $  &  $1.5\times10^{-1}$   & $\lsim V_{cd}$ & $V_{cd} = 0.23$\\
\hline
$\nu_{\tau} \tau su $  &   $6.0 \times10^{-3}$  & unitarity& $\delta V_{us} = .0027$\\
\hline
$\nu_{i}\tau su$  &   $7.7 \times10^{-2}$  & unitarity& $\delta V_{us} = .0027$\\

\hline

$\nu_{\tau}\tau sc$   & $1.2 \times10^{-1}$   & unitarity & $\delta V_{cs} = .06$\\
\hline
%
$\nu_{i} \tau sc$  & $2.4 \times10^{-1}$   & unitarity & $\delta V_{cs} = .06$\\
\hline
$\nu_{i} \tau bu$   & $3.9 \times 10^{-3}$ & $\lsim V_{ub}$ & $<3.93\times10^{-3}$\\
\hline
$\nu_{i} \tau bc$  & $4.1 \times 10^{-2}$ & $ \lsim V_{cb}$ & $<4.12\times10^{-2}$\\
\hline
\end{tabular}
\caption{
Constraints on the dimensionless
coefficient  $\epsilon^{ijkn}$, of the four-fermion
interaction $2 \sqrt{2}  G_F \left(\bar{\nu}_{i} \gamma^{\mu} P_{L}
e_{j}\right)\left(\bar{d}_{k}\gamma_{\mu} P_{L,R} u_{n}\right) $.
 The
 indices $ijkn$ are  given  in the left colomn,
the  best calculated constraints  are in
column 2, arising the  observable  of column 3, with the  
experimental value in column 4. 
}
\label{tab:CC}
\end{center}
\end{table}

\begin{table}[!tpb]
\begin{center}
\begin{tabular}{|c|c|c|c|}
\hline
&&&\\
$\left(\bar{\nu}_{i}{e}_{j}\right)\left(\bar{d}_{k}u_{n}\right)$ & Constraint on $\epsilon^{ijkn}$ & Observable & Experimental value\\
&&& \\

\hline
\hline
$\nu_{e}edu$  & $6.3 \times10^{-7}$   & $R_{\pi}$& $(1.230\pm 0.004)\times10^{-4}$\\
\hline
$\nu_{i}edu$  & $1.6 \times10^{-5}$   & $R_{\pi}$& $(1.230\pm 0.004)\times10^{-4}$\\
\hline
$\nu_{i}edc$  &   $1.4 \times10^{-3}$   & $\Gamma(D^+ \to \bar{e} \nu_i)$& 
$< 8.8 \times10^{-6}$\\  
\hline
$\nu_{e}e su$& $2.3\times10^{-6}$ & $BR(K^+ \to \bar{e} \nu_e)$ 
 & $(1.55\pm 0.07)\times10^{-5}$\\
\hline
$\nu_{i}e su$& $1.5\times10^{-5}$ & $BR(K^+ \to \bar{e} \nu_i)$ 
 & $(1.55\pm 0.07)\times10^{-5}$\\
\hline
$\nu_{i}esc$  & $ 4.9 \times 10^{-3}$     & $BR(D_s^+ \to \bar{e} \nu_i)$   
& $ < 1.3 \times 10^{-4} $ \\
\hline
$\nu_{i}ebu$   & $1.8\times10^{-4}$ & $BR(B^{+}\rightarrow \bar{e}\nu)$ & $<5.2\times10^{-6}$\\
\hline
$\nu_{i}ebc$  &     &  & \\
\hline
$\nu_{e}\mu su$    & $9.7\times10^{-4}$ & $BR(K^{+}\rightarrow \bar{\mu}\nu_{e})$ & $<4.0\times10^{-3}$\\
\hline
$\nu_{\mu}\mu du$   & $1.3 \times10^{-4}$ & $R_{\pi}$& $(1.230\pm 0.004)\times10^{-4}$\\
\hline
$\nu_{i}\mu du$   & $3.2 \times10^{-3}$ & $R_{\pi}$& $(1.230\pm 0.004)\times10^{-4}$\\
\hline
$\nu_{\mu}\mu dc$   & $7.6 \times10^{-4}$   
& $BR(D^{+}\rightarrow \bar{\mu}\nu)$ & $(3.82\pm 0.33)\times10^{-4}$\\
\hline
$\nu_{i}\mu dc$  &  $3.7 \times10^{-3}$   
& $BR(D^{+}\rightarrow \bar{\mu}\nu)$ & $(3.82\pm 0.33)\times10^{-4}$\\ 
\hline
$\nu_{\mu}\mu su$ & $2.4\times10^{-4}$ & $\frac{f_{s}}{f_{+}(0)}$ for $K^{+}_{\mu3}$ & $0.2\times10^{-2}$\\
\hline
$\nu_{i}\mu su$   & $3.0\times10^{-3}$ & $R_{K}$& $(2.44\pm 0.11)\times10^{-5}$\\
\hline
$\nu_{\mu}\mu sc$   & $4.3\times10^{-3}$   
& $\frac{BR(D^{+}_{s}\rightarrow\bar{\tau}\nu_{\tau})}{BR(D^{+}_{s}\rightarrow \bar{\mu}\nu_{\mu})}$ & $(11.0\pm 1.4 \pm0.6) $\\
\hline
$\nu_{i} \mu sc$  & 
 $1.7\times10^{-2}$   
& $\frac{BR(D^{+}_{s}\rightarrow\bar{\tau}\nu )}
{BR(D^{+}_{s}\rightarrow \bar{\mu}\nu)}$ & $(11.0\pm 1.4 \pm0.6) $\\
\hline
$\nu_{i}\mu bu$   & $1.0\times10^{-4}$  & $BR(B^{+}\rightarrow \bar{\mu}\nu)$ & $<1.7\times10^{-6}$\\
\hline
$\nu_{i} \mu bc $  &     &  & \\
\hline
$\nu_{\tau} \tau du $  &  $4.5\times10^{-3}$     &$BR(\tau \rightarrow \pi^+ \nu)$  & $ (10.91 \pm 0.07) \times 10^{-2} $\\
\hline
$\nu_{i} \tau du $  &  $8.0\times10^{-2}$     &$BR(\tau \rightarrow \pi^+ \nu)$  & $ (10.91 \pm 0.07) \times 10^{-2} $\\
\hline
$\nu_{i} \tau dc $  &  $1.5\times10^{-1}$   & $BR(D^{+}\rightarrow \bar{\tau}
\nu) $ & $<1.2 \times10^{-3}$\\
\hline
$\nu_{\tau} \tau su $ & $2.3\times10^{-2}$ & $BR(\tau \rightarrow K^+ \nu)$      
& $(6.96\pm 0.23) \times 10^{-3} $ \\
\hline
$\nu_{i} \tau su $ & $2.2\times10^{-1}$ & $BR(\tau \rightarrow K^+ \nu)$      
& $(6.96\pm 0.23) \times 10^{-3} $ \\
\hline
$\nu_{\tau}\tau sc$   & $7.2\times10^{-2}$   
& $\frac{BR(D^{+}_{s}\rightarrow\bar{\tau}\nu_{\tau})}{BR(D^{+}_{s}\rightarrow \bar{\mu}\nu_{\mu})}$ & $(11.0\pm 1.4 \pm0.6) $\\
\hline
$\nu_{i} \tau sc$  & 
 $2.9\times10^{-1}$   
& $\frac{BR(D^{+}_{s}\rightarrow\bar{\tau}\nu )}
{BR(D^{+}_{s}\rightarrow \bar{\mu}\nu)}$ & $(11.0\pm 1.4 \pm0.6) $\\
\hline
$\nu_{i} \tau bu $& $8.2\times 10^{-4}$  & $BR(B^{+}\rightarrow \bar{\tau}\nu_{\tau})$ & $(1.4 \pm 0.4) \times10^{-4}$\\ 
\hline
$\nu_{\tau}\tau bu$   & $3.1\times10^{-4}$  & $BR(B^{+}\rightarrow \bar{\tau}\nu_{\tau})$ & $(1.4 \pm 0.4) \times10^{-4}$\\
\hline
$\nu_{i} \tau bc $  &     &  & \\
\hline
\end{tabular}
\caption{Constraints  from ``charged current'' processes on
$S\pm A$ operators. These apply to  $\epsilon^{ijkn}_{\ell qS}$
and $\epsilon^{ijkn}_{ qde}$.
The first colomn is the index combination
$ijkn$,  the second is the constraints, which
 arise from the observable  given  in column 3.
The  experimental value  used is the last colomn.
$\nu_i$ is any flavour of neutrino. }
\label{tab:SpmA}
\end{center}
\end{table}

\begin{table}[!tpb]
\begin{center}
\begin{tabular}{|c|c|c|c|}
\hline
&&&\\
$\left(\bar{\ell}_{i}{e}_{j}\right)\left(\bar{q}_{k}u_{n}\right)$ & Constraint on $\epsilon^{ijkn}_{\ell qS}$ & Observable & Experimental value\\
&&& \\

\hline
\hline
$eeuc$   & $7.6\times10^{-4}$ & $BR(D^{0}\rightarrow \bar{e}e)$ & $<1.2\times10^{-6}$\\
\hline
$\mu euc$  & $6.3\times10^{-4}$& $BR(D^{0}\rightarrow \bar{\mu}e)$ & $<8.1\times10^{-7}$\\
\hline
$\mu\mu uc$   & $7.9\times10^{-4}$& $BR(D^{0}\rightarrow \bar{\mu}\mu)$ & $<1.3\times10^{-6}$\\
\hline
\end{tabular}
\caption{Constraints on $\epsilon^{ijkn}_{\ell qS}$ 
for the $ijkn$ index combination given in the first column.
The bounds given in table \ref{tab:SpmA}
also apply. 
The second colomn is the constraint, which
 arises from the observable  given  in column 3.
The  experimental value  used is the last colomn.
These bounds  are also valid under lepton  
and/or quark index permutation.}
\label{Constraints on lqS}
\end{center}
\end{table}

\begin{table}[!tpb]
\begin{center}
\begin{tabular}{|c|c|c|c|}
\hline
&&&\\
$\left(\bar{\ell}_{i}e_{j}\right)\left(\bar{d}_{k}q_{n}\right)$ & Constraint on $\epsilon^{ijkn}_{qde}$ & Observable & Experimental value\\
&&& \\
\hline
\hline
$eeds$    & $2.1\times10^{-8}$ & $BR(\overline{K^{0}_{L}}\rightarrow \bar{e}e)$ & $9.0\times10^{-12}$\\
\hline
$eedb$    & $3.2\times10^{-5}$  & $BR(\overline{B^{0}}\rightarrow \bar{e}e)$ & $<1.13\times10^{-7}$\\
\hline
$eesb$    & $5.6\times10^{-4}$ & $BR(\overline{B^{0}_{s}}\rightarrow \bar{e}e)$ & $<5.4\times10^{-5}$\\
\hline
$\mu eds$    & $9.0\times10^{-9}$   & $BR(\overline{K^{0}_{L}}\rightarrow \bar{\mu}e)$ & $<4.7\times10^{-12}$\\
\hline
$e\mu bd$  & $2.9\times10^{-5}$ & $BR(B^{0}\rightarrow \bar{e}\mu)$ & $<9.2\times10^{-8}$\\
\hline
$e\mu bs$  & $1.9\times10^{-4}$ & $BR(B^{0}_{s}\rightarrow \bar{e}\mu)$ & $<6.1\times10^{-6}$\\
\hline
$e\tau ds$   &&& \\
\hline
$e\tau db$  & $1.1\times10^{-3}$ & $BR(\overline{B^{0}}\rightarrow \bar{e}\tau)$ & $<1.1\times10^{-4}$\\
\hline
$e\tau sb$   &&& \\
\hline
$\mu\mu ds$   & $5.9\times10^{-7}$ & $BR(\overline{K^{0}_{L}}\rightarrow \bar{\mu}\mu)$ & $6.84\times10^{-9}$\\
\hline
$\mu\mu db$    & $1.2\times10^{-5}$  & $BR(\overline{B^{0}}\rightarrow \bar{\mu}\mu)$ & $<1.5\times10^{-8}$\\
\hline
$\mu\mu bs$     & $1.7\times10^{-5}$ & $BR(B^{0}_{s}\rightarrow \bar{\mu}\mu)$ & $<4.7\times10^{-8}$\\
\hline
$\mu\tau bd$  & $6.6\times10^{-4}$ & $BR(B^{0}\rightarrow \bar{\mu}\tau)$ & $<3.8\times10^{-5}$\\
\hline
$\mu\tau ds$  & &&\\
\hline
$\mu\tau sb$  & &&\\
\hline
$\tau \tau ds$ &&&\\
\hline
$\tau \tau db$    & $8.0\times10^{-3}$  & $BR(\overline{B^{0}}\rightarrow \bar{\tau}\tau)$ & $<4.1\times10^{-3}$\\
\hline
$\tau \tau sb$ &&&\\
\hline
\end{tabular}
\caption{Constraints on $\epsilon^{ijkn}_{qde}$	
for the $ijkn$ index combination given in the first column.
The bounds given in table \ref{tab:SpmA}
also apply. 
The second colomn is the constraint, which
 arises from the observable  given  in column 3.
The  experimental value  used is the last colomn.
These bounds  are also valid under lepton  
and/or quark index permutation.}
\label{Constraints on qde}
\end{center}
\end{table}

\clearpage

\section{Expectations for  flavour structure}
\label{expect}

In this section, we aim to make ``motivated'' guesses for
the  flavour structure of the two-quark, two lepton operator
coefficients.  We prefer not to use the
predictions of  Minimal Flavour Violation (MFV)
\cite{MFV,dAGIS} for two reasons:
from a phenomenological perspective, 
there is not a unique extension to the lepton sector
\cite{MFVL}, and in a more model-dependent
approach, defining MFV for New Physics such
as leptoquarks is even more ambiguous \cite{DGI}.

Instead, we consider an  alternative to MFV, 
which provides almost enough suppression of 
Flavour Changing Neutral Currents (FCNC)
in the quark sector.  Following Cheng and Sher\cite{Cheng:1987rs}, 
we assume NP couplings
have a flavour hierarchy patterned on
the Yukawa couplings : $\epsilon^{ijkn} \sim
\sqrt{y_i y_j y_k y_n}$ where $y_i$ is
the Yukawa couplings of fermion $i$.
This can 
arise in models with
extra dimensions\cite{ED},
or can  be described  in  $4$ dimensions  by  inverse hierachies 
of  the $Z$ coefficients 
of the fermion  kinetic terms (the hierarchy  
being  imprinted  on all interactions when the
kinetic terms are canonically normalised \cite{GSS}).

The Yukawa couplings of $u,c,t$ are 
$
y_{u,c,t} = \frac{g m_{u,c,t}}{2 m_W}  
$, and for charged lepton and $d$-type
quarks,   a factor of $\tan \beta$ is included:
$$
y_{d,s,b} = \tan \beta \frac{g m_{d,s,b}}{2 m_W}  ~~. 
$$
We include the rescaling parameter $\tan \beta$ to allow
for the possibility that the underlying theory
of flavour physics gives  hierarchies of
flavoured couplings. In this perspective,
the largest eigenvalue of all matrices
of flavoured couplings might be $\sim 1$, with 
the magnitude of dimensionful parameters
like masses controlled by some other physics
(such as Higgs vevs).

To obtain a flavour structure for  the coefficients
of two-lepton two quark operators, 
 we assume that the scale  of new
physics is
$$
\frac{1}{m_{NP}^2} \simeq \frac{4 G_F}{\sqrt{2}}
$$
and that the coefficient of an operator containing
$u_i$  (or $d_i$, $e_i$), will contain a factor
$\sqrt{y_{ui}}$ (or $\sqrt{y_{di}}, \sqrt{y_{ei}}$).  For the doublets,
we assume that $q_i$ comes with a factor $\sqrt{y_{ui}}$
(because its larger), and $\ell_i$ with a factor $\sqrt{y_{ei}}$
(because we do not consider neutrino masses). 
This suggests a
hierarchy of order
\beq
\epsilon_{(1)\ell q} ^{ijkn} ,~
\epsilon_{(3)\ell q} ^{ijkn} ,~
\epsilon_{e u} ^{ijkn} ,  ~
\epsilon_{\ell u} ^{ijkn} , ~
\epsilon_{\ell q S} ^{ijkn} 
\sim \sqrt{ y_{e}^i  y_{e}^j y_{u}^k  y_{u}^n}
\label{expect1}
\eeq 
\beq
\epsilon_{e d}^{ijkn} ,~
\epsilon_{\ell d}^{ijkn} ,~
\epsilon_{q d e}^{ijkn} 
 \sim   y_{d}^k  y_{d}^n 
\sqrt{\frac{ y_{e}^i  y_{e}^j}{ y_{u}^k  y_{u}^n}} 
 ~~~\longrightarrow~~~  \sqrt{ y_{e}^i  y_{e}^j y_{d}^k  y_{d}^n}
\label{expect2}
\eeq
for the coefficients of the operators listed in equations
(\ref{O1lq}) to (\ref{5}).  Since $m_b<m_t$, $m_s<m_c$ and
$m_d \sim 2 m_u$, we replace  $ (y_{d}^k  y_{d}^n )^2/ y_{u}^k  y_{u}^n 
\to  y_{d}^k  y_{d}^n $ in eqn  (\ref{expect2}), and use
the estimate to the right of the arrow. The experimental
contraints are weaker than our guesses, even with this optimistic
approximation.

In tables \ref{attentes1} and  table \ref{attentes2},
  we respectively  list, in the first colomn,  all
the flavour index combinations $e_i e_ju_ku_n$
and  $e_i e_jd_kd_n$. Then in the following colomns
we give   the best
bound we obtained for  any chiral structure,  the expected
value of $\epsilon^{ijkn}$ for $\tan \beta = 1$, and the observable from
which the bound is obtained. The bounds
which arise from charged current processes are
specifically labelled,  because  it is less
clear whether the ``expectation'' would be 
$ \propto \sqrt{ y_{e}^i  y_{e}^j y_{d}^k  y_{d}^n}$
 or $\propto \sqrt{ y_{e}^i  y_{e}^j y_{u}^k  y_{u}^n}$.

The  tables show that in most cases, for
$\tan \beta = 1$, 
our naive expectations are far below  the
experimental sensitivity  of the processes
we have considered. (Recall, however, that 
we neglected most loop processes. There would
be additional constraints, for instance
from meson-anti-meson mixing recently studied by \cite{Saha:2010vw}, 
on any new
particles inducing our operators.)
In boldface, we draw attention to 
some  kaon observables, where the 
current bounds are close to our guesses.
This suggests that more sensitive rare
Kaon experiments could be a good place
to look for two lepton, two quark operators.

\begin{table}[htp!]
\begin{center}
\begin{tabular}{|c|c|c|c|}
\hline
&&&\\
$ijkn$

 & Constraint on $\epsilon^{ijkn}$ & Expectation & Observable \\
&& &\\
\hline
\hline
$ee uu$ &    $1 \times10^{-2}$ &$ 5  \times 10^{-11 } $  & $\Lambda^{\pm}_{ee uu;LL}$ \\
$\nu_{e}e ud $ &    $6  \times10^{-7}$ &  &  $R_{\pi}$
\\
\hline
$eeuc$   &  $ 8 \times10^{-4}$    & $  1  \times 10^{-9 } $ &  $BR(D^0 \to \bar{e} e)$  \\
$\nu_{e}e us $& $2.3\times10^{-6}$& & $BR(K^+ \to \bar{e} \nu_e)$ \\
\hline
$\nu_{i}eub$   & $1.8\times10^{-4}$ &&
 $BR(B^{+}\rightarrow \bar{e}\nu)$ \\
\hline
$ee cc$ &  $1 \times10^{-2}$  &   $   2  \times 10^{-8 } $   & $\Lambda^{+}_{eecc;LL}$ \\
\hline
$ee ct$ & $ $ &   $ 2  \times 10^{-7 } $ & \\
\hline
$ee tt$ & $ 9  \times 10^{-2 } $ &   $ 3  \times 10^{-6 } $& $ Z \to \bar{e} e$ \\
\hline
$e\mu uu$ & $8.5\times10^{-7}$ & $ 6  \times 10^{-10 } $  & $\mu-e$ conversion \\
\hline
$e\mu uc$ & $ 6 \times10^{- 4}$     &  $ 1  \times 10^{-8 } $ &
 $D^0 \to \bar{\mu}e $\\
\hline
$\nu_{e}\mu ub$   & $1.0\times10^{-4}$  & $ 2  \times 10^{-7 } $  
& $BR(B^{+}\rightarrow \bar{\mu}\nu)$\\
\hline
$e\mu cc$ &  0.6 &   $ 3  \times 10^{-7 } $& $Z \to \bar{\mu}e $\\
\hline
$e\mu ct$ &  &   $ 3  \times 10^{-6 } $& \\
\hline
$e\mu tt$ & .1 & $ 4 \times 10^{-5 } $ &  $Z \to \bar{\mu}e $\\
\hline
$e \tau  uu$ & $8 \times10^{-4}$ &  $  3  \times 10^{-9 }$ &  $\frac{BR(\tau^{-}\rightarrow \pi^{0}e^{-})}{BR(\tau^{-}\rightarrow \pi^{-}\nu_{\tau})}$\\ 
$ e \nu_{\tau} ud$  & $1.6\times10^{-5}$  &  & $R_\pi$\\
\hline
$e \tau uc$ &   & $ 6  \times 10^{-8 } $ & \\
\hline
$e \tau ut$ &   & $ 8  \times 10^{-7 } $ & \\
$\nu_{i}ebu$   & $1.8\times10^{-4}$ && $BR(B^{+}\rightarrow \bar{e}\nu)$\\
\hline
$e \tau cc$ &  2     &   $  2  \times 10^{-6 } $ &  $Z \to \bar{\mu}e $\\
\hline
$e \tau ct$ &    & $  2  \times 10^{-5 }$ & \\
\hline
$e \tau tt$ & 
 0.2      &  $  2  \times 10^{-4 }$ &  $Z \to \bar{\mu}e $\\
\hline
$\mu\mu uu$  &   $ 1  \times 10^{-2 }$ & $  9  \times 10^{-9 }$
& $\Lambda^{-}_{\mu \mu uu;RR}$\\
$\nu_{\mu}\mu du$   & $1.3 \times10^{-4}$ && $R_{\pi}$\\
\hline
$\mu\mu uc$  & $8 \times10^{-4}$    & $  2  \times 10^{-7 }$ & 
$BR(D^0 \to \mu \bar{\mu})$\\
$\nu_{\mu}\mu su$ & $2.4\times10^{-4}$ && $\frac{f_{s}}{f_{+}(0)}$ for $K^{+}_{\mu3}$\\
\hline
$ \mu \mu  ut$ &   & $  2  \times 10^{-6 }$ & \\
\hline
$\mu\mu cc$  & $1  $    & $  4  \times 10^{-6 }$& $Z \to \bar{\mu} \mu$\\
$\nu_{\mu}\mu sc$   & $4.3\times10^{-3}$   &
& $\frac{BR(D^{+}_{s}\rightarrow\bar{\tau}\nu_{\tau})}{BR(D^{+}_{s}\rightarrow \bar{\mu}\nu_{\mu})}$\\
\hline
$ \mu \mu  ct$ &   & $  5  \times 10^{-5 }$ &\\
\hline
$\mu\mu tt$  & $0.07 $    & $  6  \times 10^{-4 }$ & $Z \to \bar{\mu} \mu$\\
\hline
$\mu\tau uu$ & $9.8\times10^{-4}$ & $ 3  \times 10^{-8 } $ &  $\frac{BR(\tau^{-}\rightarrow \pi^{0}e^{-})}{BR(\tau^{-}\rightarrow \pi^{-}\nu_{\tau})}$\\
\hline
$\mu \tau uc$ &  &   $  6  \times 10^{-7 } $&\\
$\nu_{i}\mu su$   & $3.0\times10^{-3}$ && $R_{K}$\\
\hline
$\mu \tau ut$ &    & $  8  \times 10^{-6 } $&\\
$\nu_{i}\mu bu$   & $1.0\times10^{-4}$  && $BR(B^{+}\rightarrow \bar{\mu}\nu)$ \\
\hline
$\mu \tau cc$ & 2 &$ 2  \times 10^{-5 }$& $Z \to \bar{\tau} \mu$\\
\hline
$\mu \tau ct$ &   & $ 2  \times 10^{-4 } $&\\
\hline
$\mu \tau tt$ & 0.07 &$  2  \times 10^{-3 }$& $Z \to \bar{\tau} \mu$\\
\hline
$\tau \tau uu$ &  0.5 & $ 2  \times 10^{-7 }$& $Z \to \bar{\tau} \tau$\\
\hline
$\tau \tau uc$ &     & $ 3  \times 10^{-6 }$&\\
\hline
$\tau \tau ut$ &   & $ 4  \times 10^{-5 }$ &\\
\hline
$\tau \tau cc$  & 0.5  & $ 7  \times 10^{-5 }$& $Z \to \bar{\tau} \tau$\\
\hline
$\tau \tau ct$  &  & $ 8  \times 10^{-4 }$ &\\
\hline
$\tau \tau tt$  &0.09  & $ 1  \times 10^{-2 }$& $Z \to \bar{\tau} \tau$\\
\hline

\end{tabular}
\caption{ This table indicates the proximity of 
 current experimental bounds to  the hierarchical
expectation:    $\epsilon^{ijkn}
\simeq \sqrt{ y_e^i y_e^j y_u^k y_u^n}$.
The first colomn  gives all
the flavour index combinations $e_i e_ju_ku_n$.
 The second   colomn  is   the best
bound we obtained for  any chiral structure, the
third colomn is   the expected
value of $\epsilon^{ijkn}$, and the last is the observable from
which the bound is obtained. Notice that 
  $\epsilon \propto \tan \beta$, 
so it is straightforward to see which observables
become interesting as $\tan \beta$ increases.
}
\label{attentes1}
\end{center}
\end{table}

\begin{table}[htp!]
\begin{center}
\begin{tabular}{|c|c|c|c|}
\hline
&&&\\
$
ijkn$
 & Constraint on $\epsilon^{ijkn}$ &
Expectation & Observable \\
&&& \\
\hline
\hline
$eedd$   &  $1.0 \times10^{-2}$ & $ 9  \times 10^{-11 }$  & $\Lambda^{\pm}_{eedd;LL}$ \\
\hline
$eeds$   & $5.7\times10^{-5}$ & $ 3  \times 10^{-10 }$&  $BR(K^{0}_{L}\rightarrow \bar{e}e)$\\
$\nu_e\nu_e ds$ & $9.4\times10^{-6}$     
&& $\frac{BR(K^{+}\rightarrow \pi^{+}\bar{\nu}\nu)}
{BR(K^{+}\rightarrow\pi^{0}\bar{e}\nu_{e})}$ \\
\hline
$eedb$   & $2.0\times10^{-4}$   & $ 3  \times 10^{-9 }$   &  $\frac{BR(B^{+}\rightarrow \pi^{+}\bar{e}e)}
{{BR(B^{0}\rightarrow\pi^{-}\bar{e}\nu_{e})}}$\\
\hline
$eess$   & $1.5 \times 10^{-2}$   & $ 2  \times 10^{-9 }$& $ \Lambda_{eess:LL}^\pm$\\
\hline
$eesb$   & $1.8\times10^{-4}$    & $ 1 \times 10^{-8}$   &  
$\frac{BR(B^{+}\rightarrow K^{+}\bar{e}e)}
{{BR(B^{0}\rightarrow D^{0}\bar{e}\nu_{e})}}$\\
\hline
$eebb$   &  $1.5\times10^{-2}$ & $8 \times10^{-8}$  & $\Lambda^{-}_{eebb;LL}$ \\
\hline
$e\mu dd$ & $8.5\times10^{-7}$ & $ 1 \times10^{-9}$ & $\mu -e $ conversion\\
\hline
{\boldmath $e\mu ds$ }   & {\boldmath  $9.0\times10^{-9}$ }   &  {\boldmath $ 4 \times 10^{-9} $} & $BR(\overline{K^{0}_{L}}\rightarrow \bar{\mu}e)$  \\
\hline
$e\mu db$ & $2.9\times10^{-5}$     
& $  4 \times 10^{-8} $& $BR({B^{0}}\rightarrow \bar{\mu}e)$\\
\hline
$e\mu ss$   & $ 0.5$    & $ 2 \times 10^{-8} $    & $Z \to \bar{e} \mu$  \\
\hline
$e\mu sb$ & $8 \times10^{-5}$       &   $ 2 \times 10^{-7} $ & 
$\frac{BR(B^{+}\rightarrow K^{+}\bar{\mu}e)}
{{BR(B^{0}\rightarrow D^{0}\bar{e}\nu_{e})}}$\\
\hline
$e\mu bb$   & $ 3$    & $ 1 \times 10^{-6} $    & $Z \to \bar{e} \mu$  \\
\hline
$e \tau  dd $ & $8.4\times10^{-4}$ & $6 \times 10^{-9}$ &  
$\frac{BR(\tau^{-}\rightarrow \pi^{0}e^{-})}
{BR(\tau^{-}\rightarrow \pi^{-}\nu_{\tau})}$ \\
\hline
$e\tau ds$ & $4.9\times10^{-4} $     & $ 2 \times 10^{-8}$& 
$\frac{BR(\tau^{-}\rightarrow K^{0}e^{-})}
{BR(\tau^{-}\rightarrow K^{-}\nu_{\tau})}$\\
$\nu_e\nu_\tau ds$ & $9.4\times10^{-6}$     
&& $\frac{BR(K^{+}\rightarrow \pi^{+}\bar{\nu}\nu)}
{BR(K^{+}\rightarrow\pi^{0}\bar{e}\nu_{e})}$ \\
\hline
$e\tau db$ & $1.1\times10^{-3}$ & $2 \times10^{-7} $    & $BR(B^0 \to \tau \bar{e})$\\
\hline
$e\tau ss$ & 1 &   $1 \times10^{-7} $    & $Z \to \bar{e} \tau$  \\
\hline
$e\tau sb$ &  &  $8 \times10^{-7} $  & $$  \\
$\nu_e\nu_\tau sb$ & $1.0 \times10^{-3}$     
&& $\frac{BR(B^{+}\rightarrow K^{+}\bar{\nu}\nu)}
{BR(B^{+}\rightarrow D^{0}\bar{e}\nu_{e})}$ \\
\hline
$e\tau bb$ &  1 &   $5 \times10^{-6} $    & $Z \to \bar{e} \tau$  \\
\hline
$\mu\mu dd$   &  $3.2 \times10^{-3}$ &$2 \times10^{-8} $ & $R_{\pi}$\\
&  $4.3\times10^{-2}$  &  & $\Lambda^{+}_{\mu\mu dd;RR}$ \\
\hline
{\boldmath  $\mu\mu ds$}   &
{\boldmath  $5.9\times10^{-7}$} & {\boldmath $6 \times10^{-8} $} & 
$BR(\overline{K^{0}_{L}}\rightarrow \bar{\mu}\mu)$ \\
\hline
$\mu\mu db$   & $1.2\times10^{-5}$  &$6 \times10^{-7} $& $BR(B^0 \to \bar{\mu} \mu)$\\
\hline
$\mu\mu ss$    &  0.8 &   $4 \times10^{-7} $    & $Z \to \bar{\mu} \mu$  \\
& $3.0\times10^{-2}$ &   & $\frac{BR(D^{+}_{s}\rightarrow \bar{\mu}\nu_{\mu})}
{BR(D^{+}_{s}\rightarrow\bar{\tau}\nu_{\tau})}$ \\
\hline
$\mu\mu sb$   & 
 $1.7\times10^{-5}$ & $2 \times10^{-6} $  & $BR(B^{0}_{s}\rightarrow \bar{\mu}\mu)$ \\
\hline 
$\mu\mu bb$   &  0.8 &   $2 \times10^{-5} $    & $Z \to \bar{\mu} \mu$  \\
\hline
$\mu\tau dd$ & $9.8\times10^{-4}$ & $6 \times10^{-8}$ & $\frac{BR(\tau^{-}\rightarrow \pi^{0}\mu^{-})}{BR(\tau^{-}\rightarrow \pi^{-}\nu_{\tau})}$\\
\hline
$\mu \tau ds$ & $5.4\times10^{-4} $    & $2 \times10^{-7}$ &  $\frac{BR( \tau \rightarrow {\mu} K)}
{BR( \tau \rightarrow \bar{\nu} K)}$\\
$\nu_\mu \nu_\tau  ds$ & $9.4\times10^{-6}$     
&& $\frac{BR(K^{+}\rightarrow \pi^{+}\bar{\nu}\nu)}
{BR(K^{+}\rightarrow\pi^{0}\bar{e}\nu_{e})}$ \\
\hline
$\mu\tau db$ & $6.6 \times10^{-4}$   & $2 \times10^{-6}$ &  $BR(B^{0}\rightarrow \bar{\mu}\tau)$ \\
\hline
$\mu \tau ss$ &  1    & $1 \times10^{-6}$ & $Z \to \tau \bar{\mu}$ \\
\hline
$\mu \tau sb$ & $ 4\times10^{-3}$   & $ 8 \times10^{-6}$
&  $\frac{BR(B^+ \to K^+ \bar{\tau} \mu)}
{{BR(B^{+}\rightarrow D^{0}\bar{\tau}\nu)}}$ \\
$\nu_\mu \nu_\tau sb$ & $1.0 \times10^{-3}$     
&& $\frac{BR(B^{+}\rightarrow K^{+}\bar{\nu}\nu)}
{BR(B^{+}\rightarrow D^{0}\bar{e}\nu_{e})}$ \\
\hline
$\mu \tau bb$ &  1    & $5 \times10^{-5}$ & $Z \to \tau \bar{\mu}$ \\
\hline
$\tau \tau dd$  &  0.8 &   $3 \times10^{-7} $    & $Z \to \bar{\tau} \tau$  \\
\hline
{\boldmath $\tau \tau ds$} &      &{\boldmath   $1 \times10^{-5} $} &  \\
$\nu_\tau \nu_\tau ds$ &{\boldmath $9.4\times10^{-6}$}     
&& $\frac{BR(K^{+}\rightarrow \pi^{+}\bar{\nu}\nu)}
{BR(K^{+}\rightarrow\pi^{0}\bar{e}\nu_{e})}$ \\
\hline
$\tau\tau db$   & $0.2 $  &  $1 \times10^{-5} $
& $BR(B^{0}\rightarrow \bar{\tau}\tau)$ \\
\hline
$\tau \tau ss$  &  0.8 &   $6 \times10^{-6} $    & $Z \to \bar{\tau} \tau$  \\
\hline
$\tau \tau sb$ & &   $4 \times10^{-5} $      & $$  \\
$\nu_\tau \nu_\tau  sb $ & $1.0 \times10^{-3}$     
&& $\frac{BR(B^{+}\rightarrow K^{+}\bar{\nu}\nu)}
{BR(B^{+}\rightarrow D^{0}\bar{e}\nu_{e})}$ \\
\hline
$\tau \tau bb$  &  0.8 &   $3 \times10^{-4} $    & $Z \to \bar{\tau} \tau$  \\
\hline
\end{tabular}
\caption{
This table indicates the proximity of 
 current experimental bounds to  the hierarchical
 ``expectation'' :    $\epsilon^{ijkn}
\simeq \sqrt{ y_e^i y_e^j y_d^k y_d^n}$
The first colomn  gives all
the flavour index combinations 
 $e_i e_jd_kd_n$. The following  colomns are   the best
bound we obtained for  any chiral structure,  the expected
value of $\epsilon^{ijkn}$ for $\tan \beta = 1$, and the observable from
which the bound is obtained. In boldface are observables
that are flavour combinations where the bound and the
expectation are close. Notice that 
  $\epsilon \propto \tan^2 \beta$, 
so it is straightforward to see which observables
become interesting as $\tan \beta$ increases.}
\label{attentes2}
\end{center}
\end{table}

\section{Conclusion}

We have compiled flavour dependent  bounds on  effective interactions
between  two leptons and  two quarks, which could 
be induced by   $SU(3)\times SU(2)\times U(1)$ 
invariant  dimension six operators.
The constraints are 
listed  in tables \ref{tab:eeddND}-\ref{Constraints on qde}, 
with  the rows labelled by 
the fermion generations. 
The bounds are set assuming that only one interaction is present at a time
(so we neglect possible cancellations).    
For each possible  combination of external leg flavours,
the strongest bound  is listed. Constraints were obtained  from
rare meson decays (leptonic or semi-leptonic), 
semi-leptonic tau decays,  
contact interactions at colliders, $Z$ decay
data from LEP1,  and $\mu -e$ conversion. 

We also discussed, in section
\ref{expect}, a naive ``guess'' for the expected flavour
structure of the operator coefficients.  The expectations
are 
always below the current experimental bounds (for
$\tan \beta = 1$), as can
be seen in tables \ref{attentes1} and \ref{attentes2}. 
Some rare Kaon decay bounds are close to these
``expectations''.

\subsection*{Acknowledgments}

We thank Gino Isidori  and Thomas Schwetz
for suggestions and useful discussions, and
MC  thanks Dr. Aldo Deandrea for his explanations and availability.

\end{document}